
\input harvmac

\Title{\vbox{\baselineskip12pt\hbox{CERN-TH/95-294, DFUPG 111-95}
\hbox{UGVA-DPT 1995/11-907}}}
{\vbox{\centerline{Gauge Theories of Josephson Junction Arrays}}}


\centerline{M. C. Diamantini\footnote{$^1$}
{Supported by I.N.F.M., Section of Perugia, Italy}}
\medskip\centerline{Theory Division, CERN, CH-1211 Geneva 23, Switzerland}
\bigskip
\centerline{P.Sodano}
\medskip\centerline{I.N.F.N. and Department of Physics, University of
Perugia}
\centerline{via A. Pascoli, I-06100 Perugia, Italy}
\bigskip
\centerline{C. A. Trugenberger\footnote{$^2$}
{Supported by a Profil 2 fellowship of the Swiss National Science Foundation}}
\medskip
\centerline{Department of Theoretical Physics, University of Geneva}
\centerline{24, quai E. Ansermet, CH-1211 Geneva 4, Switzerland}

\vskip .3in
We show that the zero-temperature physics of planar Josephson
junction arrays in the self-dual approximation
is governed by an Abelian gauge theory
with {\it periodic} mixed Chern-Simons term describing
the charge-vortex coupling. The
periodicity requires the existence of (Euclidean) topological
excitations which
determine the quantum phase structure of the model. The electric-magnetic
duality leads to a quantum phase transition between a superconductor and a
superinsulator at the self-dual point. We also discuss in this framework
the recently proposed quantum Hall phases for charges and vortices
in presence
of external offset charges and magnetic fluxes: we show how the
periodicity of
the charge-vortex coupling can lead to transitions to anyon
superconductivity
phases. We finally generalize our results to three dimensions, where the
relevant gauge theory is the so-called BF system with an antisymmetric
Kalb-Ramond gauge field.

\Date{November 1995}

\newsec{Introduction}
Gauge fields can be used to model the long distance behaviour of
several condensed matter systems \ref\froa{J. Fr\"ohlich
and U. Studer, Rev. Mod. Phys. 65 (1993) 733, J. Fr\"ohlich, R. G\"otschmann
and P. A. Marchetti, J. Phys. A: Math. Gen. 28 (1995) 1169.},
a connection which has
been particularly exploited for planar systems \ref\fwz{X.-G. Wen, Int. Jour.
Mod. Phys. B6 (1992) 1711, "Topological Orders and Edge Excitations in
FQH states", MIT preprint 95-148; A. Zee, "From Semionics to
Topological Fluids",
in "Cosmology and Elementary Particles", Proceedings of the 1991 Rio Piedras
Winter School, D. Altschuler, J. F. Nieves, J. P. de Leon and M. Ubriaco eds.,
World Scientific, Singapore (1991).} .
In a nutshell, the idea is that charge fluctuations around a given
ground state are decribed by a conserved current $j^{\mu }$, which in (2+1)
dimensions can be represented in terms of a gauge field  $B_{\mu}$ according
to $j^{\mu } \propto \epsilon ^{\mu \alpha \nu}\partial _{\alpha }B_{\nu}$.
For a wide
class of systems the effective action governing the dynamics of the charge
fluctuations is quadratic in the gauge fields $B_{\mu }$ at long distances
\froa . Clearly this effective action is also gauge invariant, reflecting the
original gauge invariance of the definition of the current $j^{\mu }$: one
obtains thus an effective gauge theory at long distances (which is not
necessarily relativistic).
The ground states of a wide class of planar condensed matter systems
\ref\fra{For a general review see: E. Fradkin, "Field Theories of Condensed
Matter Systems", Addison-Wesley, Redwood City (1991).}
can thus be classified according to the lowest derivative term appearing in
their effective
gauge theory at long distances. This way Chern-Simons terms describe
incompressible quantum fluids (quantum Hall states) and chiral spin liquids
\ref\frob{B. Blok and X.-G. Wen, Phys. Rev. B42 (1990) 8133, Phys. Rev. B42
(1990) 8145; J. Fr\"ohlich and T. Kerler, Nucl. Phys. B354 (1991)
369; J. Fr\"ohlich and A. Zee, Nucl. Phys. B364 (1991) 517, X.-G. Wen and
A. Zee, Phys. Rev. B46 (1992) 2290.} while the Maxwell term describes a
(2-dim.) superfluid (superconductor) \froa \ \fwz .

In this paper we shall investigate a further connection between Abelian
gauge theories and certain condensed matter systems, namely
{\it Josephson junction arrays}
\ref\jos{U. Eckern and A. Schmid, Phys. Rev. B39 (1989) 6441;
R. Fazio and G. Sch\"on, Phys. Rev. B43 (1991) 5307;
R. Fazio, A. van Otterlo, G. Sch\"on, H. S. J. van der Zant
and J. E. Mooij, Helv. Phys. Acta 65 (1992) 228.} .
In a recent publication \ref\dst{M. C. Diamantini, P. Sodano and C. A.
Trugenberger, Nucl. Phys. B448 (1995) 505.}
\ we studied  non-perturbative features of the (2+1)-dimensional
gauge theory with mixed Chern-Simons term
\ref\mav{A. Kovner and B. Rosenstein, Phys. Rev. B42 (1990) 4748;
G. W. Semenoff and N. Weiss, Phys. Lett. B250 (1990) 117;
N.Dorey and N. E. Mavromatos, Phys. Lett. B250 (1990) 107,
Nucl. Phys. B386 (1992) 614.}
\eqn\mcs{{\cal L}=-{1\over 4e^2}F_{\mu \nu}F^{\mu \nu } + {\kappa \over 2\pi }
A_{\mu }\epsilon ^{\mu \alpha \nu }\partial_{\alpha }B_{\nu }-{1\over 4g^2}
f_{\mu \nu }f^{\mu \nu } }
when the gauge symmetries associated with the two Abelian gauge fields
$A_{\mu }$ and $B_{\mu }$ are compact. We also pointed out the relevance
of \mcs \ to the zero-temperature physics of planar Josephson junction arrays.
Here we will derive and study this connection in detail.

After reviewing  in section 2 the basic physics of \mcs \ and our
lattice notation, we shall show in section 3 that the zero-temperature
partition function of Josephson junction arrays in the
{\it self-dual approximation}
coincides with the Euclidean partition function
of the lattice version of \mcs \ with {\it periodic} mixed Chern-Simons
coupling. This means that the two gauge fields are compact variables only
as far as their coupling is concerned. The periodicity is implemented
by two types of topological
excitations
\ref\pol{For a review see: A. M. Polyakov,
"Gauge Fields and Strings", Harwood Academic Press, Chur (1987).}
\ which constitute electric and magnetic closed loops with short-range
interactions.
The two energy scales of Josephson junction arrays, the charging energy
$E_C$ and the Josephson coupling $E_J$ are directly related to the two
massive parameters $e^2$ and $g^2$ of \mcs .

In section 4 we investigate the non-perturbative structure
of this Chern-Simons lattice gauge model. The phase structure is determined
by the 3-dimensional statistical mechanics of the topological excitations
and reflects the self-duality of the model. We find three possible phases at
zero temperature. For small $e/g$ there is a {\it superconducting} phase with
logarithmic confinement of magnetic fluxes; in this phase the original $R_A$
gauge symmetry of \mcs \ is broken down to $Z_A$ so that the full symmetry
is given by $Z_A \times R_B$. Correspondingly, one of
the two massive excitations of \mcs \ becomes massless. The dual phase is
realized for large $e/g$. In this phase we have logarithmic confinement of
electric charges and symmetry $R_A \times Z_B$, with a corresponding massless
excitation. An infinite energy (voltage) is required to
separate the charge dipoles and produce  a
current through the sample: we call this phase with infinite resistance
a {\it "superinsulator"}. Depending on the details of the lattice,
a third phase can open up between the superconductor and the superinsulator.
In this phase the topological excitations are irrelevant, the symmetry
is $R_A\times R_B$ and both excitations are massive.
The amount of energy required to produce a current through
the sample is exponentially small.
In \dst \ we called this phase the Chern-Simons phase: in presence
of dissipation it would
actually correspond to a {\it "metallic"} phase of the model \ref\gor{We thank
G. W. Semenoff for pointing this out to us.} .
The superconductor-insulator quantum phase
transition is actually observed experimentally in planar Josephson
junction arrays at very low temperatures \jos .

Recently it has been suggested that Josephson junction
arrays in presence of $n_q$ offset charges and $n_{\phi }$ external magnetic
fluxes per plaquette might have quantum Hall
phases \fra \ for either charges \ref\odi{A. A. Odintsov and
Y. V. Nazarov, Phys. Rev. B51 (1995) 1133.} \ or fluxes \ref\choi{M. Y.
Choi, Phys. Rev. B50 (1994) 10088.}
\ \ref\ste{A. Stern, Phys. Rev. B50 (1994) 10092.} , depending
on the ratios $n_q/n_{\phi }$ and $E_C/E_J$.
In section 5 we discuss these purely two-dimensional quantum Hall states in
the framework of the gauge theory representation. Specifically, we show that
they can be described by additional pure Chern-Simons terms for either one
of the two gauge fields $A_{\mu }$ or $B_{\mu }$. In this phases the charges
and vortices combine to form an {\it incompressible quantum fluid}
\ref\Lau{For a review see R. B. Laughlin, "The Incompressible Quantum Fluid",
in "The Quantum Hall Effect", R. E. Prange and S. M. Girvin eds., Springer
Verlag, New York (1990).} \ of charge-flux composites with short-range
interactions. Localized excitations are charge and flux carrying anyons
\ref\lerda{For a review see: F. Wilczek,
"Fractional Statistics and Anyon Superconductivity", World Scientific,
Singapore (1990); A. Lerda, "Anyons", Springer-Verlag, Berlin (1992).} .

We then investigate how one of the distinctive
feature of Josephson junction arrays, namely the
periodicity of charge-vortex couplings affects these quantum Hall states.
We find that this periodicty can induce two types of phase
transitions. The charge-flux fluid corresponding to the
charge quantum Hall phase can either expel the flux and form a
charge superfluid corresponding to
a conventional superconductor or condense into a charge-flux superfluid.
Correspondingly, the flux-charge fluid corresponding to the vortex
quantum Hall
phase can either expel the charge and form a flux superfluid corresponding to
a superinsulator or condense into a flux-charge superfluid.
These superfluids of charge-flux composites are
(logarithmic ) {\it oblique confinement phases}
\ref\hoo{G. 't Hooft, Nucl. Phys. B190 [FS3](1981) 455.}
\ \ref\car{J. L. Cardy and E. Rabinovici,
Nucl. Phys. B205 [FS5] (1982) 1;
J. L. Cardy, Nucl. Phys. B205 [FS5] (1982) 17.}
\ corresponding to {\it anyon superconductors}
\ref\bobl{R. B. Laughlin, Phys. Rev. Lett. 60 (1988) 2677;
see also \lerda .} .
We thus conclude that Josephson junction arrays might provide the first
explicit realization of the anyon superconductivity mechanism.

In section 6 we generalize our results to three dimensions (even if
three-dimensional Josephson junction arrays have not yet been fabricated).
In this case, one
of the two gauge fields becomes an antisymmetric Kalb-Ramond tensor gauge
field \ref\kal{M. Kalb and P. Ramond, Phys. Rev. D9 (1974) 2273; F. Lund and
T. Regge, Phys. Rev. D14 (1976) 1524.} and the
(3+1)-dimensional gauge theory we obtain
is the so-called BF-model \ref\bal{A. P. Balachandran, V. P. Nair, B.-S.
Skagerstam and A. Stern, Phys. Rev. D26 (1982) 1443; T. J. Allen, M. Bowick
and A. Lahiri, Mod. Phys. Lett. A6 (1991) 559; A. P. Balachandran and P.
Teotonio-Sobrinho, Int. Jour. Mod. Phys. A9 (1994) 1569.} . This is an Abelian
gauge model with a conventional Maxwell gauge field and a Kalb-Ramond gauge
field coupled by a topological mass term. In three dimensions the magnetic
topological excitations become compact surfaces on the lattice and
self-duality is lost. The zero-temperature phase structure is determined by
the statistical mechanics of a model of coupled random loops and random
surfaces in four Euclidean dimensions: this can also be viewed as the
Euclidean partition function for a lattice model of particles interacting with
closed Nielsen-Olesen type strings.
While the statistical mechanics of random loops is by
now well developed \pol
\ \ref\loo{See for example: C. Itzykson and J.-M. Drouffe, "Statistical
Field Theory", Cambridge University Press, Cambridge (1989).}
\ there is no corresponding amount of analytical results for random
surfaces \ref\froc{Excellent reviews are: J. Fr\"ohlich, "The Statistical
Mechanics of Surfaces", in "Applications of Field Theory to Statistical
Mechanics", L. Garrido ed., Lecture Notes in Physics Vol. 216, Springer-Verlag,
Berlin (1985); "Statistical Mechanics of Membranes and Surfaces", Jerusalem
Winter School for Theoretical Physics, Vol. 5, D. Nelson, T. Piran and
S. Weinberg eds., World Scientific, Singapore (1989), F. David, "Introduction
to the Statistical Mechanics of Random Surfaces and Membranes", in
"Two-Dimensional Quantum Gravity and Random Surfaces", Jerusalem Winter School
for Theoretical Physics, Vol. 8, D. J. Gross, T. Piran and S. Weinberg eds.,
World Scientific, Singapore (1992).} . Assuming three distinct phases as in
(2+1) dimensions, with condensation of electric loops, no condensation of
topological excitations and condensation of magnetic surfaces we can identify
the first two again with superconducting and metallic phases, respectively.
In the phase with condensation of magnetic surfaces the charge dipoles are
bound by $1/r$ potentials, which are long-range but not confining. Therefore
only a finite amount of energy is required to separate them and the system
behaves as an insulator (as opposed to a superinsulator in two dimensions).

\newsec{The lattice Chern-Simons model}
Our model \mcs \ can be rewritten in terms of the dual field strengths
\eqn\dfs{\eqalign{F^{\mu } &\equiv {1\over 2}\epsilon^{\mu \alpha \beta}
F_{\alpha \beta}\ ,\qquad \qquad F_{\mu \nu} \equiv \partial _{\mu }A_{\nu }
-\partial _{\nu }A_{\mu } \ ,\cr
f^{\mu } &\equiv {1\over 2} \epsilon ^{\mu \alpha \beta} f_{\alpha \beta}
\ ,\qquad \qquad f_{\mu \nu} \equiv \partial _{\mu }B_{\nu }-\partial _{\nu }
B_{\mu } \ ,\cr }}
as follows \footnote{$^*$}{Throughout this paper we use units such that
$c=1$ and $\hbar =1$.}
\eqn\mod{{\cal L}_{CS} = -{1\over 2e^2} \left({1\over \eta}F_0F^0
+F_iF^i\right)
+{\kappa \over 2\pi }
A_{\mu }\epsilon ^{\mu \alpha \nu } \partial _{\alpha }B_{\nu } -
{1\over 2g^2}
\left( {1\over \eta }f_0f^0 +f_if^i\right) \ .}
For later convenience we have introduced a magnetic permeability $\eta $,
equal for the two gauge fields. The coupling constants $e^2$ and $g^2$ have
dimension mass, whereas  the coefficient $\kappa $ of the mixed Chern-Simons
term is dimensionless. Note that we take $B_{\mu }$ to represent a
pseudovector
gauge field, so that the mixed Chern-Simons term does not break the discrete
symmetries of parity and time reversal.

The action corresponding to \mod \ is separately invariant under the two
Abelian gauge transformations
\eqn\agt{\eqalign{A_{\mu } &\to A_{\mu } +\partial _{\mu }\lambda \ ,\cr
B_{\mu } &\to B_{\mu } +\partial _{\mu }\omega \ ,\cr}}
with gauge groups $R_A$ and $R_B$, respectively. Moreover, the action is
also invariant under the {\it duality transformation}
\eqn\dtr{\eqalign{A_{\mu } &\leftrightarrow B_{\mu }\ ,\cr
e &\leftrightarrow g\ ,\cr}}
so that the model is {\it self-dual}.

The Lagrangian \mod \ can be easily diagonalized by the linear transformation
\eqn\lit{\eqalign{A_{\mu } &= \sqrt {e\over g} \left( a_{\mu }+b_{\mu }
\right)
\ ,\cr
B_{\mu } &= \sqrt{g\over e} \left(a_{\mu }-b_{\mu } \right) \ .\cr }}
In terms of these  new variables the model \mod \ describes a free theory,
\eqn\tla{{\cal L}_{CS}= -{1\over eg} \left( {1\over \eta }G_0G^0
+G_iG^i \right)
+{\kappa \over 2\pi }a_{\mu }\epsilon^{\mu \alpha \nu }\partial _{\alpha}
a_{\nu } - {1\over eg} \left({1\over \eta }g_0g^0 +g_ig^i\right) -
{\kappa \over 2\pi}b_{\mu }\epsilon^{\mu \alpha \nu}\partial_{\alpha }b_{\nu }
\ ,}
where $G^{\mu }$ and $g^{\mu }$ are the dual field strengths for the new gauge
fields $a_{\mu }$ and $b_{\mu }$, respectively. This Lagrangian describes
a doublet of excitations with topological mass \ref\jac{R. Jackiw and
S. Templeton, Phys. Rev. D23 (1981) 2291; J. Schonfeld, Nucl. Phys. B185
(1981) 157; S. Deser, R. Jackiw and S. Templeton, Phys. Rev. Lett. 48 (1982)
975, Ann. Phys. (N.Y.) 140 (1982) 372.}
\eqn\tma{m={|\kappa|eg\over 2\pi }\ ,}
and  spectrum
\eqn\spe{E({\bf q})=\sqrt{m^2+{1\over \eta}|{\bf q}|^2} \ .}

In the following we shall formulate a Euclidean lattice version of
the above Chern-Simons model. To this end we introduce a three-dimensional
rectangular lattice with lattice spacings $l_{\mu }$ in the three directions.
In particular we shall take the lattice spacings $l_1=l_2\equiv l$ and
identify $l_0$ with the spacing in the Euclidean time direction. Lattice
sites are denoted by the three-dimensional vector $x$; the gauge fields
$A_{\mu }(x)$ and $B_{\mu }(x)$ are associated with the links $(x, \mu )$
between the sites $x$ and $x+\hat \mu $, where $\hat \mu $ denotes a unit
vector in direction $\mu $ on the lattice.

On the lattice we introduce the following forward and backward derivatives and
shift operators:
\eqn\dso{\eqalign{d_{\mu } f(x) &\equiv {{f(x+l_{\mu }\hat \mu )-f(x)}\over
l_{\mu }}\ ,\qquad \qquad S_{\mu }f(x) \equiv f(x+l_{\mu }\hat \mu )\ ,\cr
\hat d_{\mu } f(x) &\equiv {{f(x)-f(x-l_{\mu }\hat \mu )}\over l_{\mu }} \ ,
\qquad \qquad \hat S_{\mu }f(x) \equiv f(x-l_{\mu }\hat \mu ) \ . \cr }}
Summation by parts on the lattice interchanges both the two derivatives (with
a minus sign) and the two shift operators; gauge transformations are defined
using the forward lattice derivative. Corresponding to the two derivatives
in \dso , we can define also two lattice analogues of the Chern-Simons
operators
$\epsilon _{\mu \alpha \nu }\partial _{\alpha }$
\ref\frod{J. Fr\"ohlich and P. A. Marchetti,
Comm. Math. Phys. 121 (1989) 177; P. L\"uscher,
Nucl. Phys. B326 (1989) 557; V. F. M\"uller, Z. Phys. C47 (1990) 301; D.
Eliezer and G. W. Semenoff, Ann. Phys. (N.Y.) 217 (1992) 66.} \ \dst :
\eqn\lcs{k_{\mu \nu } \equiv S_{\mu } \epsilon _{\mu \alpha \nu } d_{\alpha }
\ , \qquad \qquad \hat k_{\mu \nu } \equiv \epsilon _{\mu \alpha \nu } \hat d
_{\alpha }\hat S_{\nu } \ ,}
where no summation is implied over equal indices $\mu $ and $\nu $.
Summation by parts on the lattice interchanges also these two operators
(without an extra minus sign).
The operators \lcs \ are both local and gauge invariant, in the sense that
\eqn\gin{k_{\mu \nu} d_{\nu }=\hat d_{\mu }k_{\mu \nu }=0\ ,\qquad \qquad
\hat k_{\mu \nu }d_{\nu }=\hat d_{\mu }\hat k_{\mu \nu } =0 \ ,}
and their product reproduces the relativistic, Euclidean lattice
Maxwell operator:
\eqn\lmo{k_{\mu \alpha }\hat k_{\alpha \nu } = \hat k_{\mu \alpha }
k_{\alpha \nu } = -\delta _{\mu \nu }\nabla ^2 +d_{\mu }\hat d_{\nu } \ ,}
where $\nabla ^2 \equiv \hat d_{\mu }d_{\mu }$ is the
three-dimensional Laplace
operator. Using $k_{\mu \nu }$ we can also define the lattice dual field
strengths as
\eqn\ldfs{\eqalign{F_{\mu } &\equiv \hat k_{\mu \nu }A_{\nu }\ ,\cr
f_{\mu } &\equiv k_{\mu \nu }B_{\nu } \ .\cr}}
The identity \lmo \ then tells us that we can simply write the relativistic,
Euclidean lattice Maxwell terms as $\sum _x F_{\mu }F_{\mu }$ and
$\sum _x f_{\mu }f_{\mu }$.

Using all these definitions we can now write the Euclidean lattice partition
function of our model \mod \ as follows:
\eqn\lpf{\eqalign{Z_{CS} &= \int {\cal D}A_{\mu }
\int {\cal D}B_{\mu } \ {\rm exp}(-S_{CS}) \ ,\cr
S_{CS} &= \sum_x {l_0l^2\over 2e^2} \left( {1\over \eta }F_0F_0
+F_iF_i \right)
-i {l_0l^2\kappa \over 2\pi } A_{\mu }k_{\mu \nu }B_{\nu } + {l_0l^2 \over
2g^2} \left( {1\over \eta }f_0f_0 +f_if_i \right) \ ,\cr }}
where we have introduced the notation ${\cal D}A_{\mu }\equiv
\prod _{(x, \mu )} dA_{\mu }(x)$ and gauge fixing is understood.

For later convenience we introduce also the finite difference operators
\eqn\fdo{\Delta _{\mu } \equiv l_{\mu }d_{\mu }\ , \qquad \qquad
\hat \Delta _{\mu } \equiv l_{\mu }\hat d_{\mu } \ ,}
where no summation over equal indices is implied. Correspondingly, we
introduce also the finite difference analogue of the operators $k_{\mu \nu }$
and $\hat k_{\mu \nu }$:
\eqn\kbig{K_{\mu \nu } \equiv S_{\mu }\epsilon _{\mu \alpha \nu }
\Delta _{\alpha } \ ,\qquad \qquad \hat K_{\mu \nu } \equiv \epsilon_{\mu
\alpha \nu }\hat \Delta_{\alpha }\hat S_{\nu } \ .}
These satisfy equations analogous to \gin \ and \lmo \ with all derivatives
substituted by finite differences.

\newsec{Josephson junction arrays}
Josephson junction arrays \jos \ are quadratic, planar arrays of spacing
$l$ of superconducting islands with nearest neighbours Josephson couplings
of strength $E_J$. Each island has a capacitance $C_0$ to the ground;
moreover there are also nearest neighbours capacitances $C$. The Hamiltonian
characterizing such systems is thus given by
\eqn\hjja{H=\sum_{\bf x} \ {C_0\over 2} V_{\bf x} + \sum _{<{\bf x \bf y}>}
\left( {C\over 2} \left( V_{\bf y}-V_{\bf x} \right) ^2 + E_J
\left( 1-{\rm cos}\ N\left( \Phi _{\bf y} - \Phi _{\bf x} \right) \right)
\right) \ ,}
where boldface characters denote the sites of the two-dimensional array,
$<{\bf x \bf y}>$ indicates nearest neighbours, $V_{\bf x}$ is the electric
potential of the island at ${\bf x}$ and $\Phi _{\bf x}$ the phase of its
order parameter.  For generality we allow for any integer $N$ in the Josephson
coupling, so that the phase has periodicity $2\pi /N$: obviously $N=2$ for the
real systems.

With the notation introduced in the previous section the
Hamiltonian \hjja \ can be rewritten as
\eqn\hjjb{H= \sum_{\bf x} \ {1\over 2} V \left( C_0 - C\Delta \right) V +
\sum _{{\bf x},i} \ E_J \left( 1-{\rm cos} \ N \left( \Delta_i \Phi \right)
\right) \ ,}
where $\Delta \equiv \hat \Delta_i \Delta_i$ is the two-dimensional finite
difference Laplacian and we have omitted the explicit location indeces on
the variables $V$ and $\Phi$.

The phases $\Phi _{\bf x}$ are quantum-mechanically conjugated to the charges
$Q_{\bf x}$ on the islands: these are quantized in integer multiples of $N$
(Cooper pairs for $N=2$):
\eqn\cqu{Q= q_e N p_0\ ,\qquad \qquad p_0 \in Z\ ,}
where $q_e$ is the electron charge. The Hamiltonian \hjjb \ can be expressed
in terms of charges and phases by noting that the electric
potentials $V_{\bf x}$ are determined by the charges $Q_{\bf x}$ via
a discrete version of Poisson's equation:
\eqn\dvp{\left( C_0 -C\Delta \right) V_{\bf x} = Q_{\bf x} \ .}
Using this in \hjjb \  we get
\eqn\hjjc{H= \sum_{\bf x} \ N^2E_C \ p_0 {1\over {C_0\over C}-\Delta } p_0 +
\sum _{{\bf x},i} \ E_J \left( 1-{\rm cos} \ N\left( \Delta _i \Phi \right)
\right) \ ,}
where $E_C\equiv q_e^2/2C$. The integer charges $p_0$ interact via a
two-dimensional Yukawa potential of mass $\sqrt{C_0/C} /l$. In the
nearest-neighbours capacitance limit $C\gg C_0$, which is accessible
experimentally, this becomes essentially a two-dimensional Coulomb law.
{}From now on we shall consider the limiting case $C_0=0$. In this case the
{\it charging energy} $E_C$ and the {\it Josephson coupling} $E_J$ are the
two relevant energy scales in the problem. These two massive parameters can
also be traded for one massive parameter $\sqrt{2N^2E_CE_J}$, which represents
the {\it Josephson plasma frequency} and one massless parameter $E_J/E_C$.

The zero-temperature partition function of the Josephson junction array
admits a (phase-space) path-integral representation \ref\nor{For a review see:
J, W. Negele and H. Orland, "Quantum Many-Particle Systems", Addison-Wesley,
Redwood City (1988).} . Since the variables $p_0$ are integers, the
imaginary-time integration has to be performed stepwise; we
introduce therefore
a lattice spacing $l_0$ also in the imaginary-time direction. This has to be
just smaller of the typical time scale
on which the integers $p_0$ vary, in
the present case the inverse of the Josephson plasma frequency:
$l_0 \le O\left( 1/\sqrt{2N^2E_CE_J} \right) $. We thus get the following
partition function:
\eqn\pfjja{\eqalign{Z &= \sum_{\{p_0\}} \int_{-\pi /N}^{+\pi /N} {\cal D}\Phi
\ {\rm exp}(-S)\ ,\cr
S &= \sum _x -iN\ p_0\Delta _0\Phi + N^2 E_C l_0 \ p_0 {1\over -\Delta } p_0
+\sum _{x, i} \ l_0E_J \left( 1- {\rm cos}\ N \left( \Delta _i \Phi \right)
\right) \ ,\cr }}
where now the sum in the action $S$ extends over the three-dimensional lattice
with spacing $l_0$ in the imaginary time direction and $l$ in the spatial
directions.

In the next step we introduce vortex degrees of freedom by replacing the
Josephson term by its Villain form \ref\kle{For a review see: H. Kleinert,
"Gauge Fields in Condensed Matter", World Scientific, Singapore (1989).} :
\eqn\pfjjb{\eqalign{Z &= \sum _{{\{p_0\} \atop \{v_i\}}}
\int _{-\pi /N}^{+\pi /N} {\cal D}\Phi \ {\rm exp}(-S)\ ,\cr
S &= \sum_x -iN\ p_0\Delta _0 \Phi + N^2 E_C l_0 \ p_0 {1\over -\Delta } p_0
+N^2 l_0 {E_J\over 2} \left( \Delta _i \Phi + {2\pi \over N} v_i \right) ^2
\ .\cr }}
Strictly speaking, this substitution is valid only for $l_0 E_J \gg1$; however
the Villain approximation retains all most relevant features of the
Josephson coupling for the whole range of values of the coupling
$E_J$ \kle \ and therefore we shall henceforth adopt it.

We now represent the Villain term as a Gaussian integral over real variables
$p_i$ and we transform also $p_0$ to a real variable by introducing new
integers $v_0$ via the Poisson summation formula
\eqn\poi{\sum_{k=-\infty }^{k=+\infty } {\rm exp}(i2\pi kz)= \sum_{n=-\infty}
^{n=+\infty } \delta (z-n) \ .}
By grouping together the real and integer $p$ and $v$ variables into
three-vectors $p_{\mu }$ and $v_{\mu }$, $\mu=0,1,2$ we can write the
partition function as
\eqn\pfjjc{\eqalign{Z &= \sum _{\{ v_{\mu } \} } \int
{\cal D}p_{\mu } \int_{-\pi /N}^{+\pi /N} {\cal D}\Phi \ {\rm exp}(-S)\ ,\cr
S &= \sum_x -iN p_{\mu } \left( \Delta _{\mu } \Phi + {2\pi \over N} v_{\mu }
\right) +N^2 E_C l_0 \ p_0 {1\over -\Delta} p_0 +
{p_i^2\over 2l_0E_J} \ .\cr }}

Following \pol \ we use the longitudinal part of the integer vector field
$v_{\mu }$ to shift the integration domain of $\Phi$. To this end we
decompose $v_{\mu }$ as follows:
\eqn\dif{v_{\mu }= \Delta _{\mu } m + \Delta _{\mu }\alpha + K_{\mu \nu}
\psi _{\nu }\ ,}
where $m\in Z$, $|\alpha |<1$ and $K_{\mu \nu }$ defined in \kbig . Here the
vectors $\psi _{\mu }$ are not integer, but they are nonetheless restricted
by the fact that the combinations $q_{\mu } \equiv \hat K_{\mu \nu }v_{\nu }=
\hat K_{\mu \alpha }K_{\alpha \nu } \psi_{\nu }$ must be integers. The
original sum over the three independent integers $\{ v_{\mu } \}$ can thus
be traded for a sum over the four integers $\{ m, q_{\mu } \}$ subject to
the constraint $\hat \Delta _{\mu }q_{\mu } =0$. The sum over the integers
$\{ m \}$ can then be used to shift the $\Phi$ integration domain from
$[-\pi /N, +\pi /N )$ to $(-\infty, +\infty )$. The integration over $\Phi$
is now trivial and enforces the constraint $\hat \Delta _{\mu }p_{\mu } =0$:
\eqn\pfjjd{\eqalign{Z &= \sum _{\{ q_{\mu } \}} \delta_{\hat \Delta_{\mu }
q_{\mu }, 0} \ \int {\cal D}p_{\mu } \ \delta \left(
\hat \Delta _{\mu }p_{\mu } \right) \ {\rm exp }(-S)\ ,\cr
S &= \sum_x -i2\pi \ p_{\mu } K_{\mu \nu } \psi_{\nu } + N^2 l_0 E_C \ p_0
{1\over -\Delta } p_0 + {p_i^2 \over 2l_0E_J} \ .\cr }}

We now solve the two constraints by introducing a real gauge field $b_{\mu }$
and an integer gauge field $a_{\mu }$:
\eqn\sco{\eqalign{ p_{\mu } &\equiv K_{\mu \nu }b_{\nu } \ , \qquad
\qquad b_{\mu }\in R\ ,\cr
q_{\mu } &\equiv \hat K_{\mu \nu } a_{\nu }\ ,\qquad
\qquad a_{\mu }\in Z\ .\cr }}
By inserting the first of these two equations and by summing by parts, the
first term in the action \pfjjd \ reduces to $\sum_x -i2\pi b_{\mu }q_{\mu }$.
By inserting the second of the above equations and by summing by parts
again, this term of the action finally reduces to the mixed Chern-Simons
coupling $\sum_x -i2\pi \ a_{\mu }K_{\mu \nu }b_{\nu }$. Using the Poisson
formula \poi \ we can finally make $a_{\mu }$ also real at the expense of
introducing a set of integer link variables $\{ Q_{\mu }\}$ satisfying
the constraint $\hat \Delta_{\mu }Q_{\mu }$, which guarantees
gauge invariance:
\eqn\pfjje{\eqalign{Z &= \sum _{\{ Q_{\mu } \}} \int {\cal D}a_{\mu }
\int {\cal D}b_{\mu } \ {\rm exp}(-S) \ ,\cr
S &= \sum_x -i2\pi \ a_{\mu }K_{\mu \nu }b_{\nu } + N^2l_0E_C \ p_0 {1\over
-\Delta }p_0 + {p_i^2\over 2l_0E_J} + i2\pi a_{\mu }Q_{\mu } \ .\cr }}
In this representation $K_{\mu \nu }b_{\nu }$ represents the conserved
three-current of charges, while $\hat K_{\mu \nu }a_{\nu }$ represents
the conserved three-current of vortices. Note that, actually, both these
conserved currents are integers (the factors of $N$ are explicit): indeed, the
summation over $\{ Q_{\mu } \}$ makes $a_{\mu }$ (and
therefore also $\hat K_{\mu \nu }a_{\nu }$) an integer, and then
the summation over $\{ a_{\mu }
\}$ makes $K_{\mu \nu}b_{\nu }$ an integer. The third term in the action
\pfjje \ contains two parts: the longitudinal
part $\left( p_i^L \right) ^2$ describes the
Josephson currents and represents a kinetic term for the charges;
the transverse part $\left( p_i^T \right) ^2$ can be rewritten as
a Coulomb interaction term for the vortex density $q_0$ by solving
the Gauss law enforced by the Lagrange multiplier $b_0$.

The partition function \pfjje \ displays
a high degree of symmetry between the charge and the vortex
degrees of freedom.
The only term which breaks this symmetry (apart from the integers $Q_{\mu }$)
is encoded in the kinetic term for the charges (Josephson currents).
This near-duality between charges and vortices has already been often invoked
in the literature \jos \ to explain the experimental quantum phase diagram
at very low temperatures. Here we introduce what we call the
{\it self-dual approximation} of Josephson junction arrays. This consists
in adding to the action in \pfjje \ a bare kinetic term for the vortices
\footnote{$^*$}{Note that such a kinetic term is anyhow induced by
integrating out the charge degrees of freedom.} and combining this with
the Coulomb term for the charges into $\sum_x {\pi ^2 \over N^2 l_0
E_C} q_i^2$. The coefficient is chosen so that the transverse
part of this term reproduces exactly the Coulomb term for the charges
upon solving the Gauss law enforced by the Lagrange multiplier $a_0$. The
longitudinal part, instead, represents the additional bare kinetic term
for the vortices. Given that now the gauge field $a_{\mu }$ has acquired
a kinetic term, we are also forced to introduce new integers $M_{\mu }$
via the Poisson formula to guarantee that the charge current $K_{\mu \nu }
b_{\nu }$ remains an integer:
\eqn\sda{\eqalign{Z_{SD} &= \sum_{\{ Q_{\mu } \} \atop \{ M_{\mu } \} }
\int {\cal D}a_{\mu } \int {\cal D}b_{\mu } \ {\rm exp}(-S_{SD})\ ,\cr
S_{SD} &= \sum_x -i2\pi \ a_{\mu }K_{\mu \nu }b_{\nu } + {p_i^2\over 2l_0E_J}
+{\pi ^2 q_i^2 \over N^2l_0E_C} + i2\pi a_{\mu }Q_{\mu } +i 2\pi b_{\mu }
M_{\mu } \ ,\cr }}
where the new integers satisfy the constraint $\hat \Delta _{\mu }M_{\mu }=0$
to guarantee gauge invariance.
After a rescaling
\eqn\res{\eqalign{A_0 &\equiv {2\pi \over \sqrt{N}l_0}a_0\ ,\qquad \qquad
A_i \equiv {2\pi \over \sqrt{N}l} a_i \ ,\cr
B_0 &\equiv {2\pi \over \sqrt{N}l_0} b_0 \ ,\qquad \qquad
B_i \equiv {2\pi \over \sqrt{N}l} b_i \ ,\cr }}
we obtain finally
\eqn\sdb{\eqalign{Z_{SD} &= \sum_{ \{ Q_{\mu } \} \atop \{ M_{\mu } \} }
\int {\cal D}A_{\mu } \int {\cal D}B_{\mu } \ {\rm exp} (-S_{SD}) \ ,\cr
S_{SD} &= \sum_x {l_0 l^2\over 2e^2} \ F_iF_i -i {l_0l^2 \kappa \over 2\pi }
A_{\mu }k_{\mu \nu }B_{\nu } + {l_0l^2 \over 2g^2} \ f_if_i \cr
&\ \ \ \ \ \ \ \ + i\sqrt{\kappa}
\left( l_0Q_0A_0 + lQ_iA_i \right) +i \sqrt{\kappa} \left( l_0M_0B_0 +
lM_iB_i \right) \ ,\cr }}
where $F_i$ and $f_i$ are defined in \ldfs \ and
\eqn\ide{e^2 = 2N E_C\ ,\qquad \kappa = N \ ,
\qquad g^2 = {4\pi ^2 \over N }E_J \ .}
This is exactly the partition function of our lattice Chern-Simons model
\lpf \ in the limit of infinite magnetic permeability $\eta = \infty $ and
with additional, integer-valued link variables $Q_{\mu }$ and $M_{\mu }$
coupled to the two gauge fields. Note that, with the above identifications,
the topological Chern-Simons mass \tma \ coincides with the Josephson plasma
frequency:
\eqn\tmjpf{m=\sqrt{2N^2 E_CE_J} \ .}
In the physical case $N=2$ this reduces to $m=\sqrt{8E_CE_J}$.
{}From the kinetic terms in \sdb \ we can also read off the charge
and vortex masses:
\eqn\cvma{\eqalign{m_q &= {1\over l^2 g^2} = {N\over 4\pi ^2 l^2 E_J} \ ,\cr
m_{\phi } &= {1\over l^2e^2} = {1\over 2Nl^2E_C} \ .\cr }}

In the regime $ml\le O(1)$, which is typically experimentally relevant, we can
choose $l_0=l$: in this case the infinite magnetic permeability constitutes
the only non-relativistic effect in the physics of Josephson junction arrays
in the self-dual approximation. However, we expect this
non-relativistic effect
to be irrelevant as far as the phase structure and the charge-vorticity
assignements are concerned. Therefore, for simplicity, we shall henceforth
consider the relativistic model, by setting $l_0=l$ and $\eta =1$, although
it is not hard to incorporate a generic value of $\eta $ into our
subsequent formalism:
\eqn\frm{\eqalign{Z_{SD} &= \sum _{ \{ Q_{\mu } \} \atop \{ M_{\mu } \} }
\int {\cal D}A_{\mu } \int {\cal D}B_{\mu } \ {\rm exp} (-S_{SD}) \ ,\cr
S_{SD} &= \sum_x {l^3\over 2e^2} F_{\mu }F_{\mu } -i{l^3\kappa \over 2\pi }
A_{\mu }k_{\mu \nu }B_{\nu } + {l^3\over 2g^2} f_{\mu }f_{\mu }
+il\sqrt{\kappa } A_{\mu }Q_{\mu } +il\sqrt{\kappa }
B_{\mu }M_{\mu } \ .\cr }}
Josephson junction arrays in the self-dual approximation constitute thus a
further, experimentally accessible example of the ideas presented in \froa
\ and \fwz . The action in \frm \ provides in fact a pure gauge theory
representation of a model of interacting charges and vortices, represented
by the conserved currents
\eqn\cav{\eqalign{q_{\mu }^{\rm charge} &\equiv {{\kappa }^{3\over 2}\over
2\pi } \ k_{\mu \nu }B_{\nu } \ ,\cr
\phi _{\mu }^{\rm vortex} &\equiv {1\over 2\pi {\kappa}^{1\over 2}}
\ \hat k_{\mu \nu }A_{\nu } \ ,}}
where the prefactors are chosen so that the quantum of charge is given by
$\kappa $, while the quantum of vorticity is given by $1/\kappa $ (factors
of $q_e$ and $2\pi $ are absorbed in the definitions of the gauge fields and
the coupling constants).

In this framework, the mixed Chern-Simons term represents both the Lorentz
force caused by vortices on charges (coupling of $q_{\mu }^{\rm charge}$ to
the "electric" gauge field $A_{\mu }$) and, by a summation by parts,
the Magnus force \ref\mag{See for
example: V. L. Streeter, "Fluid Dynamics", McGraw-Hill, New York (1948);
P. D. McCormack and L. Crane, "Physical Fluid Dynamics", Academic Press,
New York (1973).} \ caused by charges on vortices (coupling of $\phi _{\mu }
^{\rm vortex}$ to the "magnetic" gauge field $B_{\mu }$).
The integer-valued link variables $Q_{\mu }$ and $M_{\mu }$ represent
the (Euclidean) {\it topological excitations} \pol \ in the model.
They satisfy the constraints
\eqn\const{\eqalign{\hat d_{\mu } Q_{\mu } &=0\ ,\cr
\hat d_{\mu } M_{\mu } &= 0\ .\cr }}
In a dilute phase they constitute
closed electric ($Q_{\mu }$) and magnetic ($M_{\mu }$) loops
on the lattice; in a dense phase there is the additional possibility of
infinitely long strings. Due to the constraints \const \ we can choose to
represent these topological excitations as
\eqn\frep{\eqalign{Q_{\mu } &\equiv lk_{\mu \nu }Y_{\nu }\ ,
\qquad \qquad Y_{\nu }\in Z\ ,\cr
M_{\mu } &\equiv l\hat k_{\mu \nu } X_{\nu } \ ,\qquad \qquad
X_{\mu } \in Z\ ,\cr }}
and reabsorb them in the mixed Chern-Simons term as follows:
\eqn\reab{S_{SD}=\sum _x {\dots } -i{l^3\kappa \over 2\pi} \left( A_{\mu }
-{2\pi \over l\sqrt{\kappa }} X_{\mu } \right) k_{\mu \nu } \left( B_{\nu }
- {2\pi \over l\sqrt{\kappa }} Y_{\mu } \right) + {\dots } \ .}
In this representation it is clear that the topological excitations render
the charge-vortex coupling {\it periodic} under the shifts
\eqn\shi{\eqalign{A_{\mu } &\to A_{\mu } + {2\pi \over l\sqrt{\kappa }}
\ a_{\mu }\ ,\qquad \qquad a_{\mu } \in Z\ ,\cr
B_{\mu } &\to B_{\mu } + {2\pi \over l\sqrt{\kappa }} \ b_{\mu }\ ,\qquad
\qquad b_{\mu } \in Z\ .\cr }}
In physical terms, the topological excitations implement the well-known \fra
\ periodicity of the charge dynamics under the addition of an integer
multiple of the flux quantum $1/\kappa $ per plaquette and
the (less-known) periodicity of the vortex dynamics under the addition of
an integer multiple of the charge quantum $\kappa $ per site.

If we would require that the full action (including charge-charge and
vortex-vortex interactions) \frm \ be periodic under the shifts \shi , then
we would obtain the compact Chern-Simons model studied in \dst . In this case
the relevant topological excitations would be essentially $iX_{\mu }$
and $iY_{\mu }$: since these can also describe finite open strings, there
is the additional possibility of electric and magnetic monopoles \pol .
As we showed in \dst , these monopoles play a crucial role in
the regime $ml\ll 1$.

\newsec{Phase structure analysis}
In this section we investigate symmetry aspects and non-perturbative features
of the model \frm \ due to the periodicity of the charge-vortex interactions
encoded in the mixed Chern-Simons term. As expected, these depend entirely on
the topological excitations which enforce the periodicity.

Upon a Gaussian integration the partition function \frm \ factorizes
readily as
\eqn\fact{ Z_{SD} = Z_{CS} \cdot Z_{\rm Top} \ ,}
where $Z_{CS}$ is the pure gauge part defined in \lpf \ and
\eqn\top{\eqalign{Z_{\rm Top} &= \sum_{ \{ Q_{\mu } \} \atop \{ M_{\mu } \} }
\ {\rm exp}\left( -S_{\rm Top} \right) \ ,\cr
S_{\rm Top} &= \sum_x {e^2\kappa \over 2l} \ Q_{\mu }{\delta _{\mu \nu }
\over {m^2- \nabla ^2 }} Q_{\nu } + {g^2 \kappa \over 2l} \ M_{\mu }
{\delta _{\mu \nu } \over {m^2-\nabla ^2}} M_{\nu } \cr
&\ \ \ \ \ \ \ \ +i {2\pi m^2 \over l} \ Q_{\mu } {k_{\mu \nu }\over
{\nabla ^2
\left( m^2- \nabla ^2 \right) }} M_{\nu } \ ,\cr }}
with $m$ defined in \tma , describes the contribution due to the topological
excitations. The phase structure of our model is thus determined by the
statistical mechanics of a coupled gas of closed or infinitely long electric
and magnetic strings with short-range Yukawa interactions. The scale $(1/m)$
represents the width of these strings. In our case it is of the order of the
lattice spacing $l$. The third term
in the action \top , describing the topological Aharonov-Bohm interaction of
electric and magnetic strings, vanishes for strings separated by
distances much
bigger than $(1/m)$: in this case the denominator reduces to
$m^2\nabla ^2$ and, by using either one of the two equations in \frep \ and
the constraints
\const \ one recognizes immediately
that the whole term in the action  reduces to ($i2\pi {\rm integer}$),
which is
equivalent to $0$ \footnote{$^*$}{This reflects the fact that the original
charges and vortices satisfy the Dirac quantization condition.}.

\subsec{Free energy arguments}
In order to establish the phase diagram of our model we use the free energy
arguments for strings introduced in \ref\kog{T. Banks, R. Myerson and
J. Kogut, Nucl. Phys. B129 (1977) 493; see also B. Svetitski, Phys. Rep.
132 (1986) 1.}
\ and extensively used for the
analysis of four-dimensional self-dual models \car .

The usual argument for strings with Coulomb interactions \kog
\ is that interactions between strings are unimportant
for the phase structure because small strings interact via short-range dipole
interactions, while large strings have most of their multipole
moments canceled
by fluctuations. This argument is even stronger in our case, where the
interaction is anyway short-range. Therefore one retains only the self-energy
of strings, which is proportional to their length, and phase transitions from
dilute to dense phases appear when the entropy of large strings, also
proportional to their length, overwhelms the self-energy.
We shall also neglect the interaction term between electric and magnetic
strings (imaginary term in the action \top ). This is clearly a good
approximation if both types of topological excitations are dilute.

Thus, one assigns a free energy
\eqn\free{F = \left( {le^2\kappa \over 2}G(ml) \ Q^2 + {lg^2\kappa \over 2}
G(ml) M^2 - \mu  \right) \ N }
to a string of length $L=lN$ carrying electric and magnetic quantum numbers
$Q$ and $M$, respectively. Here $G(ml)$ is the diagonal element of the lattice
kernel $G(x-y)$ representing the inverse of the operator $l^2 \left( m^2 -
\nabla ^2 \right) $. Clearly $G(ml)$ is a function of the dimensionless
parameter $ml$. The last term in \free \ represents the entropy of the string:
the parameter $\mu $ is given roughly by $\mu = {\rm ln} 5$, since at each
step the string can choose between 5 different directions. In \free \ we have
neglected all subdominant functions of $N$, like a ${\rm ln } N$
correction to the entropy.

The condition for condensation of topological excitations is obtained by
minimizing the free energy \free \ as a function of $N$. If the coefficient
of $N$ in \free \ is positive, the minimum is obtained for $N=0$ and
topological excitations are suppressed. If, instead, the same coefficient
is negative, the minimum is obtained for $N= \infty $ and the system will
favour the formation of large closed loops and infinitely long strings.
Topological excitations with quantum numbers $Q$ and $M$ condense therefore
if
\eqn\cco{{le^2\kappa G(ml)\over 2\mu } \ Q^2 +
{lg^2 \kappa G(ml)\over 2\mu } \ M^2 < 1\ .}
If two or more condensations are allowed by this condition one has to choose
the one with the lowest free energy.

The condition \cco \ describes the interior of an ellipse with semi-axes
$2\mu /(le^2\kappa G(ml))$ and $2\mu /(lg^2\kappa G(ml))$ on a square lattice
of integer electric and magnetic charges. The phase diagram is obtained by
investigating which points of the integer lattice lie inside the ellipse
as its semi-axes are varied. We find it convenient to present the results in
terms of the dimensionless parameters $lm$ and $e/g$:
\eqn\phc{\eqalign{{ml G(ml) \pi \over \mu } < 1 &\to \cases{{e\over g}<1\ ,
& electric condensation\ ,\cr {e\over g} >1\ , & magnetic condensation \ ,\cr }
\cr
{ml G(ml) \pi \over \mu } > 1 &\to  \cases{{e\over g} <
{\mu \over ml G(ml) \pi} \ , & electric condensation\ ,\cr
{\mu \over ml G(ml) \pi }<{e\over g}<{ml G(ml) \pi \over \mu } \ ,
& no condensation\ ,\cr
{e\over g} > {ml G(ml) \pi \over \mu }\ , & magnetic condensation \ .\cr }
\cr }}
As expected, these condensation patterns are symmetric around the
the point $e/g =1$, reflecting the self-duality of the model.
In first approximation the electric (magnetic) condensation phase is
characterized by the fact that $\{ Q_{\mu } \}$ ($\{ M_{\mu } \}$) fluctuate
freely, while all $M_{\mu }=0$ ($Q_{\mu }=0$). Within this approximation it is
clearly consistent to neglect altogether the interaction term between
electric and magnetic strings in \free . Taking into account small loop
corrections \kle \ in the various phases can lead to a renormalization of
coupling constants and masses and, correspondingly, to a shift of the
critical couplings $(ml)_{\rm crit}$ and $(e/g)_{\rm crit}$
for the phase transitions. A notable exception is the case
in which there is only one phase transition: in this case the
critical coupling
is $(e/g)_{\rm crit}=1$ due to self-duality.

\subsec{Wilson and 't Hooft loops}
In order to distinguish the various phases we introduce the typical
order parameters of lattice gauge theories \pol \ref\kogut{See for example
J. Kogut, Rev. Mod. Phys. 55 (1983) 775.}, namely the {\it Wilson loop} for
an electric charge $q$ and the {\it 't Hooft loop} for a vortex $\phi $:
\eqn\opa{\eqalign{L_W &\equiv {\rm exp} \left( i {q\over {\kappa }^{1\over 2}}
\sum_x lq_{\mu }A_{\mu } \right) \ ,\cr
L_H &\equiv {\rm exp} \left( i \phi {\kappa }^{3\over 2}
\sum_x l\phi _{\mu }B_{\mu }\right) \ ,\cr }}
where $q_{\mu }$ and $\phi _{\mu }$ vanish everywhere but on the links of the
closed loops, where they take the value 1. Since the loops are closed
they satisfy
\eqn\cloo{\hat d_{\mu }q_{\mu }=\hat d_{\mu }\phi _{\mu } =0 \ .}

The expectation values $\langle L_W \rangle $ and $\langle L_H \rangle $ can
be used to characterize the various phases. First of all they
measure the interaction potential between static, external
test charges $q$ and $-q$ and vortices $\phi $ and $-\phi $, respectively
\pol . Secondly, by representing the closed loops $q_{\mu }$ and $\phi _{\mu }$
as
\eqn\lst{\eqalign{q_{\mu } &\equiv lk_{\mu \nu }A_{\nu }^q\ ,\cr
\phi _{\mu } &\equiv l \hat k_{\mu \nu }A_{\nu }^{\phi } \ ,\cr }}
we can rewrite the Wilson and 't Hooft loops as
\eqn\whl{\eqalign{L_W &= {\rm exp} \left( i{q\over {\kappa }^{1\over 2}} \sum_x
l^2 A_{\mu }^q F_{\mu } \right) \ ,\cr
L_H &= {\rm exp} \left( i {\kappa }^{3\over 2} \phi  \sum_x
l^2 A_{\mu }^{\phi }f_{\mu } \right) \ ,\cr }}
which is a lattice version of Stoke's theorem, the integers $A_{\mu }^q$ and
$A_{\mu }^{\phi }$ ($=0,\pm 1$) representing the area elements of the surfaces
spanned by the closed loops.
The second terms of the expansions of $\langle L_W \rangle $ and $\langle
L_H \rangle $ in powers of $q$ and $\phi $ measure therefore the gauge
invariant correlation functions $\langle F_{\mu }(x) F_{\nu }(y) \rangle $ and
$\langle f_{\mu }(x) f_{\nu }(y) \rangle $.
Third, if we represent $\phi _{\mu }$ as
\eqn\nint{\phi \phi _{\mu } \equiv {l^2 \over 2\pi } \ \hat k_{\mu \nu }
A_{\nu }^{\rm e.m.}\ ,}
we can also rewrite the 't Hooft loop as
\eqn\elc{L_H = {\rm exp} \left( i\sum_x l^3
A_{\mu }^{\rm e.m.}q_{\mu }^{\rm charge} \right) \ .}
With the interpretation of $A_{\mu }^{\rm e.m.}$ as an
external electromagnetic
gauge potential the expectation value of the 't Hooft loop measures the
{\it electromagnetic response} of the system in the various phases.
An analogous relation clearly holds for the Wilson loop.

The expectation values of the Wilson and 't Hooft loops are easily
obtained by combining the definitions \opa \ with \frm :
\eqn\evwh{\eqalign{\langle L_W \rangle &= {{Z_{\rm Top} \left( Q_{\mu } +
{q\over \kappa } q_{\mu }, M_{\mu } \right) } \over {Z_{\rm Top} \left(
Q_{\mu }, M_{\mu } \right) }} \ ,\cr
\langle L_H \rangle &= {{Z_{\rm Top } \left( Q_{\mu }, M_{\mu }+ \phi \kappa
\phi _{\mu } \right) } \over {Z_{\rm Top} \left( Q_{\mu }, M_{\mu } \right)
}} \ ,\cr }}
where the notation is self-explanatory. In the following we shall analyze
these expressions in the various phases obtained in \phc . We shall mostly
only indicate the form of small loop corrections: a full renormalization
group analysis is beyond the scope of the present paper and we won't be
able to predict the orders of the phase transitions.

Let us begin with the {\it electric condensation phase}. In this phase the
ground state contains many infinitely long electric strings $Q_{\mu }$.
These have a crucial effect on the gauge symmetry associated with the gauge
field $A_{\mu }$. To see this let us consider a gauge transformation
$A_{\mu } \to A_{\mu }+ d_{\mu } \Lambda $, where, for simplicity, we take
$\Lambda $ as a function of the component $x^1$ only. If we choose the usual
boundary conditions $F_{\mu }=f_{\mu }=0$ at infinity, the change of
the action
\frm \ under the above gauge transformation is given by
\eqn\cha{\Delta S_{SD}= \sum_{x^0, x^2} i\sqrt{\kappa } \left(
\Lambda (x^1=+\infty ) Q_1(x^1=+\infty )- \Lambda (x^1=-\infty ) Q_1(x^1=
-\infty ) \right) \ .}
In a dilute phase, with only small closed loops, $Q_1(x^1=+\infty )=Q_1(x^1=
-\infty ) =0$ and the action is automatically gauge invariant. In a dense
phase, with many infinitely long strings, $Q_1(x^1=+\infty )$ and $Q_1(x^1=-
\infty )$ are generically different from zero. Gauge invariance requires that
$\Delta S_{SD}$ vanishes modulo $i2\pi $. In the dense phase this is realized
only if $\Lambda $ takes the values
\eqn\asb{\Lambda = {2\pi \over \sqrt{\kappa }} n\ , \qquad \qquad n\in Z\  ,}
at infinity. This means that, in the electric condensation phase, the
{\it global} gauge symmetry is spontaneously broken down to the discrete
gauge group
$Z$, so that the total (global) symmetry of this phase is $Z_A \times R_B$.

The Wilson loop expectation value takes a particularly simple form if the
external test charges are multiples of the charge quantum: $q=n\kappa $, $n
\in Z$. In fact, since we sum over $\{ Q_{\mu } \} $, the integer loop
variables $nq_{\mu }$ can be absorbed by a redefinition of the appropriate
$Q_{\mu }$'s, with the result
\eqn\pscr{\langle L_W(q=n\kappa ) \rangle = 1\ .}
This indicates that, in this phase, external test charges $q=n\kappa $ are
perfectly {\it screened} by the topological excitations and behave
thus freely.
In order to compute the Wilson loop expectation value for generic $q$
we have to
perform explicitly the sum over $\{ Q_{\mu } \}$. To this end we have to
remember the constraint $\hat d_{\mu }Q_{\mu }=0$. We solve this constraint
by representing $Q_{\mu }=lk_{\mu \nu }n_{\nu }$ and summing over $\{ n_{\mu }
\} $, with the appropriate gauge fixing. We then use Poisson's formula \poi
\ to turn this sum into an integral, by introducing a new set of integer
link variables $\{ k_{\mu } \}$ satisfying $\hat d_{\mu }k_{\mu }=0$ in order
to guarantee the gauge invariance under $n_{\mu }\to n_{\mu }+ld_{\mu } i$.
At this point we can perform explicitly the Gaussian integration over
$\{ n_{\mu } \}$. In the approximation of neglecting terms proportional to
$\nabla ^2/m^2$ (keeping such terms would not alter substantially the result)
the new integers $\{ k_{\mu } \}$ can be absorbed by a redefinition of the
magnetic topological excitations $\{ M_{\mu } \}$, giving the result:
\eqn\corrwil{\eqalign{\langle L_W \rangle &= {{Z_{\rm corr.} \left( q_{\mu }
\right) }\over {Z_{\rm corr.}\left( q_{\mu }=0 \right)}} \ ,\cr
Z_{\rm corr.}\left( q_{\mu } \right) &= \sum _{\{ M_{\mu } \} {\rm loops}}
{\rm exp} \sum_x \left( -{g^2\kappa \over 2l} \ M_{\mu } {{\delta _{\mu \nu }}
\over -\nabla ^2} M_{\nu } +i 2\pi {q\over \kappa} A^q_{\mu }M_{\mu } \right)
\ .\cr }}
Since the magnetic topological excitations are in a dilute phase we have
to sum only over small closed loops: in this phase the dominant part
of ${\rm ln} \langle L_W \rangle$ vanishes for generic $q$ and the whole
result is given by small loop corrections. These are identical in form
to the small loop corrections for the correlation functions in the
low-temperature phase of the three-dimensional XY model \kle ; correspondingly
the Wilson loop expectation value can be computed by exactly the same
low-temperature expansion used for the XY model \kle . The first-order term
in this expansion is obtained by considering only the smallest possible
lattice loops and gives the result
\eqn\perwillo{\langle L_W \rangle = {\rm exp} \left( 2 {\rm e}^{-{g^2 \kappa l
\over 6}} \sum_{x, \mu } \left[ {\rm cos}
\left( 2\pi {q\over \kappa } q_{\mu }
\right) -1\right] \right) \ .}
The periodicity of this result is a direct consequence of the spontaneous
symmetry breaking $R_A \to Z_A$. This implies also that the gauge invariant
correlation function reduces to
\eqn\ecpapcf{\langle F_{\mu }(x) F_{\nu }(y) \rangle \propto
\left( \delta _{\mu \nu } \nabla ^2 - d_{\mu }\hat d_{\nu } \right)
\ {\delta _{x,y}\over l^3} \ ,}
which is essentially a contact term on the scale of the lattice spacing.

The computation of the 't Hooft loop expectation value follows exactly the
same lines as the above computation of the Wilson loop. The results is
\eqn\hla{\eqalign{\langle L_H \rangle &= {\rm exp}
\left( -{g^2\kappa ^3 \phi ^2\over 2l}
\sum_x \phi_{\mu } {\delta _{\mu \nu }\over -\nabla ^2}\phi_{\nu } \right)
\ {{Z_{\rm corr}\left( \phi_{\mu }\right) }\over Z_{\rm corr} \left(
\phi_{\mu }=0\right) }\ ,\cr
Z_{\rm corr}\left( \phi_{\mu } \right) &= \sum_{\{ M_{\mu } \} {\rm loops}}
{\rm exp} \left( -{g^2\kappa \over 2l} \sum_x  M_{\mu }
{\delta _{\mu \nu } \over -\nabla ^2} M_{\nu }
+2\kappa \phi \ M_{\mu } {\delta _{\mu \nu } \over -\nabla ^2}
\phi_{\nu } \right) \ .\cr }}
The first few terms in the expansion of the small loop corrections can again
be computed with the same techniques as in the low-temperature phase of the
XY model \kle .
One finds that their contribution amounts to perturbative corrections of
the Coulomb coupling constant $g^2\kappa ^3\phi ^2/2l$ of the
dominant term in \hla .

{}From \hla \ we can extract the nature of the electric condensation phase.
First of all, by considering, as usual, a rectangular loop of length $T$ in
the imaginary time direction and of length $R$ in one of the spatial
directions and computing the dominant large-$T$ behaviour of ${\rm ln}
\langle L_H\rangle $ we find that the interaction potential between external
test vortices of strength $\phi $ and $-\phi $ is proportional to ${\rm ln}R$.
Vortices are thus logarithmically {\it confined}, which amounts to the
{\it Meissner effect}. Secondly, by using the representations \lst \ and
\whl , we find the correlation function
\eqn\lrcf{\langle f_{\mu }(x) f_{\nu }(y) \rangle \propto
{{\delta _{\mu \nu } \nabla ^2 -d_{\mu }\hat d_{\nu }} \over \nabla ^2}
\ {{\delta_{x,y}}\over l^3}\ ,}
which is long-range, indicating that the "$B_{\mu }$-photon"
is {\it massless}.
This is the massless excitation associated with the spontaneous symmetry
breaking of the global gauge symmetry $R_A \to Z_A$.
Third, by using the representations \nint \ and \elc , we find that the
induced electromagnetic current is given by
\eqn\iemc{J_{\mu }^{\rm e.m.} \propto \left( \delta_{\mu \nu }-
{{d_{\mu }\hat d_{\nu }}\over  \nabla ^2} \right) \ A_{\nu }^{\rm e.m.}\ ,}
which is the standard {\it London form}. We thus conclude that the electric
condensation phase is actually a {\it superconducting phase}.

No further computation is needed to extract the nature of the {\it magnetic
condensation phase}: this is the exact dual of the electric condensation
phase just described. Specifically, the global gauge symmetry associated
with $B_{\mu }$ is spontaneoulsy broken down to $Z_B$, so that the total
symmetry of this phase is $R_A \times Z_B$. Correspondingly, the
"$A_{\mu }$-photon" is {\it massless} and the $\langle F_{\mu }(x)F_{\nu }
(y) \rangle $ correlation function is long-range. Electric charges are
logarithmically
{\it confined}, which means that an infinite energy (voltage) is required
to separate a neutral pair of charges. We call this phase with infinite
resistance a {\it superinsulator}. In real Josephson junction arrays we
expect however the conduction gap to be large but finite due to the small
ground capacity $C_0$, resulting in a normal insulator.

If $mlG(ml)\pi /\mu >1$ a third phase can open up between the superconducting
and superinsulating phases. In this third phase both the electric and the
magnetic topological excitations are dilute. Far away from the
phase transitions and to first approximation we can neglect them altogether.
This gives the result
\eqn\whmp{\eqalign{\langle L_W \rangle &= {\rm exp}
\left( -{e^2q^2\over 2l\kappa } \ q_{\mu }
{{\delta _{\mu \nu }}\over {m^2-\nabla ^2}}q_{\nu } \right) \ ,\cr
\langle L_H \rangle &= {\rm exp} \left( -{g^2 \phi ^2 \kappa ^3\over 2l}
\ \phi_{\mu } {{\delta _{\mu \nu }}\over {m^2-\nabla ^2}}
\phi_{\nu } \right) \ .\cr }}
Small loop corrections to these results can be obtained by restricting the
$\{ Q_{\mu } \}$ and $\{ M_{\mu } \}$ sums in \evwh \ to small closed loops
and using again the same techniques as in the low-temperature expansion
of the XY model \kle . These will lead to perturbative corrections of the
coupling constants and masses in \whmp ; however the first-order result
\whmp \ is enough to establish the nature of this phase. The global
symmetry characterizing this phase is $R_A\times R_B$ and, corrrespondingly,
both "photons" are massive, resulting in short-range correlation
functions $\langle F_{\mu }(x)F_{\nu }(y)\rangle$ and $\langle f_{\mu }(x)
f_{\nu }(y)\rangle $. Both charges and vortices interact via short-range
Yukawa potentials and behave thus freely when separated by distances larger
then the scale $(1/m)$. In presence of any dissipation mechanism (which would
not alter the other two phases) this third phase corresponds thus to
a {\it metallic} phase of the Josephson junction array \gor .

In conclusion we can represent the phase diagram of our model as follows:
\eqn\fphc{\eqalign{{ml G(ml) \pi \over \mu } < 1 &\to \cases{{e\over g}<1\ ,
& superconductor ($Z_A\times R_B$)\ ,\cr {e\over g} >1\ ,
& superinsulator ($R_A\times Z_B$)\ ,\cr } \cr
{ml G(ml) \pi \over \mu } > 1 &\to
\cases{{e\over g} < {\mu \over ml G(ml) \pi} \ ,
& superconductor ($Z_A\times R_B$)\ ,\cr
{\mu \over ml G(ml) \pi }<{e\over g}<{ml G(ml) \pi \over \mu } \ ,
& metal ($R_A\times R_B$)\ ,\cr
{e\over g} > {ml G(ml) \pi \over \mu }\ , & superinsulator ($R_A\times Z_B$)
\ ,\cr } \cr }}
where we have indicated in parenthesis the global symmetries of the various
phases. In \fig\met{The function ${mlG(ml)\pi \over \mu }$ for $\mu = {\rm ln}
5$.} \ we plot the (numerically computed) function $mlG(ml)\pi /\mu $ for the
value $\mu ={\rm ln}5$. This gives an indication that a window for
the metallic
phase is open for $ml$ just larger than 1, while in the regime $ml \le O(1)$,
relevant for Josephson junction arrays, a single phase transition from a
superconductor to a superinsulator at $(e/g) =1$ is favoured.

The experimental results for Josephson junction arrays are plotted in
\fig\exr{Phase diagram of fabricated Josephson junction arrays (adapted from
the last paper in \jos ). Solid squares
denote a transition from metallic behaviour to superconducting behaviour
when the temperature is lowered; open squares denote a corresponding
transition from metallic to insulating behaviour.} . These are essentially
resistance measurements as a function of temperature in arrays with
O($10^4$) cells. The zero-temperature extrapolation of these results indicates
a quantum phase transition between an insulator and a superconductor
in the vicinity of the self-dual point $E_J/E_C= 2/\pi ^2 \simeq 0.2$.

\newsec{Quantum Hall phases and anyon superconductivity}
Recently it has been suggested that, in presence of
$n_q$ offset charge quanta per site and $n_{\phi }$ external magnetic
flux quanta per plaquette in specific ratios,
Josephson junction arrays might have incompressible quantum fluid
\Lau \ phases corresponding to purely two-dimensional
{\it quantum Hall phases}
for either charges \odi \ or vortices \choi \ \ste .

In analogy with the conventional quantum Hall setting \fra \ one expects the
charge and vortex transport properties to depend on the {\it filling
fractions} $(n_q/n_{\phi})$ and $(n_{\phi}/n_q)$, respectively. Due to the
{\it periodicity} of the charge-vortex coupling, however, $n_{\phi }$ ($n_q$)
is defined only modulo an integer as far as charge (vortex) transport
properties are concerned. Using this freedom one can thus define
{\it effective filling fractions} (we shall assume
$n_q \ge 0$, $n_{\phi}\ge 0$
for simplicity):
\eqn\eff{\eqalign{\nu _q &\equiv
{n_q \over {n_{\phi }-\left[ n_{\phi } \right] ^- +
\left[ n_q \right] ^+ }} \ , \qquad \qquad 0\le \nu_q\le 1\ ,\cr
\nu_{\phi } &\equiv {n_{\phi } \over {n_q-\left[ n_q \right] ^- +\left[
n_{\phi } \right] ^+}} \ ,\qquad \qquad 0\le \nu_{\phi }\le 1 \ ,\cr }}
where $\left[ n_q \right] ^{\pm }$ indicate the smallest (greatest) integer
greater (smaller) than $n_q$. These effective filling fractions are always
smaller than 1.

The masses of charges and vortices are given in \cvma . For given values
of $n_q$ and $n_{\phi }$ which admit an incompressible quantum fluid ground
state one expects charges to bind vortices \fra \ and form a
charge quantum Hall phase in the regime where charges are
heavier than vortices, i.e.
$e/g >1$. For given values of $n_q$ and $n_{\phi }$ we thus expect a
charge quantum Hall phase at filling $\nu _q$ for $e/g >1$ and a vortex
quantum Hall phase at filling $\nu _{\phi }$ for $e/g <1$. Correspondingly,
these two regimes were analyzed in \odi \ and
\choi , respectively. In \ste , however, it was pointed out that $e/g$
cannot be too small for the vortex quantum Hall phase, since for $e/g \ll 1$
the effective vortex band mass due to the periodic array
becomes exponentially
large and vortices loose their mobility.

In the following we shall assume the existence of these quantum Hall phases
and discuss them in the framework
of the gauge theory representation of Josephson junction arrays
in the self-dual approximation. The idea is as follows.
For $n_q=n_{\phi }=0$ we
have derived that the gauge theory describing Josephson junction arrays
(in the self-dual approximation) is given by \mcs \ with periodic
charge-vortex (mixed Chern-Simons) coupling and the identifications \cav .
This gauge theory describes the dynamics of charge and vortex fluctuations
of the array in absence of external offset charges and fluxes.
Drawing on previous
experience \fwz \ \frob \ with the quantum Hall effect
we shall modify this gauge theory in order to describe charge and vortex
fluctuations about a {\it homogeneous ground state} with $n_q$ charges
and $n_{\phi }$ vortices per plaquette, describing
a quantum Hall fluid for either charges or vortices. We shall
then analyze how the periodicity of the new Chern-Simons
charge-vortex couplings affects this picture. To this end we shall consider
the Euclidean partition function of the new gauge theories,
enforcing the periodicity by appropriate topological
excitations and we shall study the resulting zero-temperature phase diagram.
Given the expected jump in the relevant effective filling fraction at
$e/g=1$, we shall consider two different gauge theories in the regimes
$e/g >1$ and $e/g <1$.

\subsec{Gauge theories for the quantum Hall phases}
Let us begin with the charge quantum Hall phase for $e/g >1$. To this end
we consider the (Minkowski space-time) gauge theory with Lagrangian
\eqn\csag{{\cal L}_q = -{1\over 2e^2} F_{\mu }F^{\mu } + {\kappa \over \pi}
A_{\mu }\epsilon ^{\mu \alpha \nu }\partial _{\alpha }B_{\nu } - {1\over
2g^2} f_{\mu }f^{\mu } - {\nu _q \over g^2}F_{\mu }f^{\mu } + {\kappa \nu _q
\over \pi }A_{\mu }\epsilon ^{\mu \alpha \nu }\partial _{\alpha }A_{\nu } \ .}
The main differences with respect to \mcs \ are the addition of a {\it pure
Chern-Simons term} for the $A_{\mu }$ gauge field and a new coupling term
proportional to $F_{\mu }f^{\mu }$. The Gauss law constraint associated
with the $A_{\mu }$ gauge field now assigns a vorticity
\eqn\voq{\phi = - {1\over 2\nu _q \kappa ^2} q }
to a charge $q=\int d^2 {\bf x} \ q_0^{\rm charge}$ (since all the gauge
fields are massive there are no corrections to this equation from boundary
terms).
The $F_{\mu }f^{\mu }$ coupling then associates a corresponding
magnetic moment
$\mu \propto \nu _q/\kappa g^2$ to these composites.
We have also rescaled the coefficient
of the mixed Chern-Simons coupling by a factor of 2
(compare with \mcs ) while maintaining the definitions \cav .
This factor of 2 is a well-known
aspect of Chern-Simons gauge theories \ref\wil{A. S. Goldhaber, R. Mackenzie
and F. Wilczek, Mod. Phys. Lett. A4 (1989) 21.} .
Indeed, the vorticity \voq \ has a back-reaction
on the charges since it also couples
to $A_{\mu }$ via the pure Chern-Simons term. With our rescaling, the total
current coupling to $A_{\mu }$ is given by $\left( 2q_{\mu }^{\rm charge} +
2\kappa ^2 \nu _q \phi _{\mu }^{\rm vortex} \right) $ and using \voq \ we see
that the total "dressed" charge of the charge-vortex composite is indeed $q$.
The rescaling ensures thus that dressed charges maintain their nominal value.

The effective Lagrangian for the charge degrees of freedom, obtained by
integrating out $A_{\mu }$ is given by
\eqn\effb{{\cal L}^B_{\rm eff.}= -{\kappa \over 4\pi \nu _q} \ B_{\mu }
\epsilon ^{\mu \alpha \nu } \partial _{\alpha }B_{\nu } + {\dots } \ ,}
where the ellipse stands for higher-derivative terms which are suppressed at
long distances by inverse powers of a mass.
Following \frob \ we introduce as external probes
a conserved vortex current $\phi _{\mu }$ and
the electromagnetic gauge field $A_{\mu }^{\rm e.m.}$:
\eqn\emco{{\cal L}^B_{\rm eff.}+A_{\mu }^{\rm e.m.} q^{\mu }_{\rm charge}
+ {\kappa }^{3\over 2} B_{\mu } \phi ^{\mu } =
{\cal L}^B_{\rm eff.} + {{\kappa ^{3\over 2}} \over 2\pi }
\ A_{\mu }^{\rm e.m.} \epsilon ^{\mu \alpha \nu }
\partial _{\alpha }B_{\nu } + {\kappa }^{3\over 2} B_{\mu } \phi ^{\mu } \ .}
Integrating also over the charge gauge field $B_{\mu }$ we find the effective
Lagrangian
\eqn\qqhp{{\cal L}_{\rm eff.}\left( A_{\mu }^{\rm e.m.}, \phi_{\mu } \right) =
{\kappa ^2 \nu _q \over 4\pi } \ A_{\mu }^{\rm e.m.}
\epsilon ^{\mu \alpha \nu }
\partial _{\alpha }A_{\nu }^{\rm e.m.} + \kappa ^2 \nu _q
\ A_{\mu }^{\rm e.m.}
\phi ^{\mu } + \pi \kappa ^2 \nu _q \ \phi_{\mu }
\epsilon ^{\mu \alpha \nu }
{\partial _{\alpha }\over \partial ^2} \phi _{\nu } \ .}
{}From this effective Lagrangian we learn two things. First of all, the
electromagnetic response of the system is encoded in the induced current
\eqn\qcurr{\eqalign{J^{\mu }_{\rm ind.} &\equiv
{\delta \over \delta A_{\mu }^{\rm e.m.}} S_{\rm eff.}
\left( A_{\mu }^{\rm e.m.}, \phi_{\mu }=0 \right)
= {\kappa ^2 \over 2\pi } \nu_q \
\epsilon ^{\mu \alpha \nu }\partial _{\alpha }A_{\nu } ^{\rm e.m.} \ ,\cr
J^i_{\rm ind.} &= -{\kappa ^2 \over 2\pi } \nu _q
\ \epsilon ^{ij}E^j \ ,\cr }}
where ${\bf E}$ is the applied electric field. This represents a {\it Hall
current} with Hall conductivity given by
\eqn\qcond{\sigma _H = {\kappa ^2 \over 2\pi} \nu _q \ .}
Secondly, the last two terms in \qqhp \ tell us that $\phi _{\mu }$ represent
charge and flux carrying anyons \lerda \ with charge-flux
relation and fractional statistics given by
\eqn\qanyo{\eqalign{q &=\nu _q \kappa ^2 \phi \ ,\cr
\theta &= \nu _q \kappa ^2 \phi ^2 \ .\cr }}
An excitation carrying no effective vorticity can be obtained by combining
a charge $q$ with a vortex $\phi = +q/2\nu _q \kappa ^2$, so that the
bare and induced vorticities cancel. This excitation has the standard
electromagnetic coupling $A_{\mu }^{\rm e.m.}q^{\mu }$, with $q^{\mu }$
representing its conserved current. The Gauss law following from \qqhp
\ then assigns to this excitation also a magnetic flux $q/\nu _q \kappa ^2$.
All excitations in the model are therefore anyons satisfying \qanyo .
Note that the magnetic moment can be written as $\mu \propto 2S/\kappa g^2$,
where $S=\nu _q/2$ is the fractional spin associated with the
fractional statistics \qanyo .

For $\nu _q = p/n$, with $p$ and $n$ coprime, $n$ flux quanta $1/\kappa $
have $p$ units $\kappa $ of charge. Since in Josephson junction arrays the
charge degrees of freedom are bosons (Cooper pairs), this excitation must
also have bosonic statistics, i.e. $\theta = {\rm even\ integer}$. This
requires that $pn$ must be an even integer. The allowed filling fractions
are thus
\eqn\qffra{q={p\over n}\ , \qquad \qquad pn = {\rm even \ integer} \ ,}
in accordance with \ref\read{N. Read, Phys. Rev. Lett. 65 (1990) 1502.} .

Note that there is no Chern-Simons term in the effective action for the
vortices, obtained by integrating out $B_{\mu }$. Indeed, the
bare and induced
Chern-Simons terms for $A_{\mu }$ cancel exactly. We thus conclude
that \csag
\ is indeed the appropriate gauge theory to describe the charge
quantum Hall
phase of Josephson junction arrays.

The gauge theory \csag \ has a {\it hidden duality}, which can
be made manifest
by rewriting the Lagrangian as
\eqn\hidu{\eqalign{{\cal L}_q &= -{1\over 2e'^2} F_{\mu }F^{\mu } +
{\kappa \over \pi}
A_{\mu }\epsilon ^{\mu \alpha \nu } \partial _{\alpha }\left( B_{\nu }+
\nu _q A_{\nu } \right) -{1\over 2g^2}
\left( f_{\mu }+\nu _q F_{\mu } \right)
\left( f^{\mu }+\nu _q F^{\mu } \right) \ ,\cr
e' &\equiv {e\over \sqrt{1-{e^2\over g^2}\nu _q^2}} \ ,\cr }}
and defining a new gauge field $B_{\mu }^q \equiv B_{\mu }+ \nu _q A_{\mu }$.
Indeed, in terms of $A_{\mu }$ and $B_{\mu }^q$, \hidu \  coincides with
\mcs \ upon substituting $e\to e'$ (and $\kappa \to 2\kappa $). In the sector
in which $f_{\mu }+ \nu _q F_{\mu } =0$ the only kinetic term of \hidu \ is
contained in $-(1/2e'^2) F_{\mu }F^{\mu }$. Therefore $m_{q\phi } \equiv
1/l^2e'^2 $ is the mass of the anyonic charge-flux composites. The gap
for collective oscillations is given by the modified topological
Chern-Simons mass
\eqn\masq{M_q \equiv m(e', g, 2\kappa )= {e'g\kappa \over \pi}
={eg\kappa \over \pi \sqrt{1-{e^2\over g^2}\nu _q^2}} \ .}
In the representation \hidu \ it is also manifest that our modified gauge
theory can be defined only in the range
\eqn\qregi{1< {e\over g} < {1\over \nu _q} \ .}
For $e/g \to 1/\nu _q$ the anyon mass $m_{q\phi }$ vanishes, while
the topological mass $M_q$ diverges.

The gauge theory describing the vortex quantum Hall phase is the dual of
\csag ,
\eqn\vgth{{\cal L}_{\phi } = -{1\over 2e^2} F_{\mu }F^{\mu } +{\kappa \over
\pi } A_{\mu }\epsilon ^{\mu \alpha \nu } \partial _{\alpha }B_{\nu } -
{1\over 2g^2} f_{\mu }f^{\mu } -{\nu _{\phi }\over e^2} F_{\mu }f^{\mu }
+{\kappa \nu_{\phi } \over \pi } B_{\mu }\epsilon ^{\mu \alpha \nu }
\partial _{\alpha }B_{\nu } \ ,}
and contains a pure Chern-Simons term for the gauge field $B_{\mu }$ and
the corresponding magnetic moment interaction.
Again, the rescaling of the mixed Chern-Simons coupling by a factor of 2
ensures that dressed vorticity maintains its nominal value.

In this case there is no pure Chern-Simons term in the effective action
for the charges, while the vortex effective Lagrangian is given by
\eqn\effa{{\cal L}^A_{\rm eff.} = -{\kappa \over 4\pi \nu _{\phi }}
A_{\mu }\epsilon ^{\mu \alpha \nu }\partial _{\alpha }A_{\nu } + \dots \ ,}
at long distances. We probe the vortex response by coupling \effa \ to
a gauge field $G_{\mu }^{\rm ext.}$ such that $\epsilon ^{\mu \alpha \nu }
\partial _{\alpha }G_{\nu }^{\rm ext.} = q^{\mu }_{\rm ext.}$
describes an external current distribution. We also introduce an additional
conserved charge current $q_{\mu }$ coupling to $A_{\mu }$ in order to
probe the quantum numbers of excitations:
\eqn\extxu{{\cal L}^A_{\rm eff.} +
{1\over \kappa ^{1\over 2}} G_{\mu }^{\rm ext.}
\epsilon ^{\mu \alpha \nu }\partial _{\alpha }A_{\nu } +
{1\over {{\kappa }^{1\over 2}}} A_{\mu }q^{\mu } \ .}
Integrating over the vortex gauge field $A_{\mu }$ we find the effective
Lagrangian
\eqn\vqhp{{\cal L}_{\rm eff.}\left( G_{\mu }^{\rm ext.}, q_{\mu } \right) =
{\pi \nu _{\phi } \over \kappa ^2} \ G_{\mu }^{\rm ext.}
\epsilon ^{\mu \alpha \nu } \partial _{\alpha }G_{\nu }^{\rm ext.} +
{2\pi \nu _{\phi } \over \kappa ^2}
\ G_{\mu }^{\rm ext.}q^{\mu } + {\pi \nu _{\phi } \over \kappa ^2} \ q_{\mu }
\epsilon ^{\mu \alpha \nu }{\partial _{\alpha } \over \partial ^2} q_{\nu }
\ .}
This leads to the following induced vortex current:
\eqn\ficu{\Phi ^{\mu }_{\rm ind.} \equiv {1\over 2\pi } {\delta \over
\delta G_{\mu }^{\rm ext.}} S_{\rm eff.} \left( G_{\mu }^{\rm ext.},
q_{\mu } =0 \right)
= {\nu _{\phi }\over \kappa ^2} \ q^{\mu }_{\rm ext.} \ .}
This equation embodies the quantum Hall effect for vortices. While charges
react to external electromagnetic fields, vortices react to external
electric currents: in the normal case the induced vortex current is
perpendicular to the applied electric current, in the quantum Hall phase it
is parallel, with coefficient proportional to $\nu _{\phi }$.
{}From the last two terms in \vqhp \ we read off the flux-charge relation and
the fractional statistics of the anyon excitations represented by $q_{\mu }$:
\eqn\fianyo{\eqalign{\phi &= {\nu _{\phi }q \over \kappa ^2} \ ,\cr
\theta &= {\nu _{\phi } q^2 \over \kappa ^2} \ .\cr }}
For $\nu _{\phi }=p/n$, $n$ charge quanta $\kappa $ carry $p$ quanta
$1/\kappa $ of flux. Since vortices are also bosons, we find the same
allowed values
of $\nu _{\phi }$ as for the charge quantum Hall phase:
\eqn\fiffra{\nu _{\phi } = {p\over n} \ , \qquad
\qquad pn={\rm even \ integer}\ .}

The self-dual representation analogous to \hidu \ is given by
\eqn\fihidu{\eqalign{{\cal L}_{\phi } &= -{1\over 2e^2}
\left( F_{\mu } + \nu _{\phi } f_{\mu } \right)
\left( F^{\mu } +\nu _{\phi } f^{\mu } \right)
+ {\kappa \over \pi } \left( A_{\mu }+\nu _{\phi } B_{\mu } \right)
\epsilon ^{\mu \alpha \nu }\partial _{\alpha } B_{\nu } -{1\over 2g'^2}
f_{\mu }f^{\mu } \ ,\cr
g' &\equiv {g\over \sqrt{1-{g^2\over e^2} \nu_{\phi }^2}}\ ,\cr }}
and coincides with \mcs \ upon introducing a new gauge field
$A_{\mu }^{\phi }
\equiv A_{\mu }+\nu _{\phi }B_{\mu }$ and substituting $g\to g'$ (and $\kappa
\to 2\kappa $). In this case $m_{\phi q} \equiv 1/l^2g'^2$ is the mass of
anyonic flux-charge composites, while the gap for collective oscillations
is given by the topological Chern-Simons mass
\eqn\masfi{M_{\phi } \equiv m(e, g', 2\kappa ) = {eg'\kappa \over \pi }
={eg\kappa \over \pi \sqrt{1-{g^2\over e^2} \nu _{\phi }^2}} \ .}
Clearly, \fihidu \ is defined only in the range
\eqn\firegi{\nu_{\phi } < {e\over g} <1\ .}
Again, for $e/g \to \nu_{\phi }$ the anyon mass $m_{\phi q}$ vanishes,
while the topological mass $M_{\phi }$ diverges.

Combining \qregi \ and \firegi \ we find the overall condition
\eqn\tora{\nu _{\phi }< e/g <1/\nu_q \ ,}
which we interpret as the regime in which a homogeneous ground state
with $n_q$ charges and $n_{\phi }$ vortices per plaquette can exist.
Presumably, for $e/g < \nu _{\phi }$ and $e/g > 1/\nu _q$ the ground state
consists of an Abrikosov-type cristal for charge-flux composites. In
particular, \tora \ tells us that in Josephson junction arrays
$E_C/E_J$ cannot be either too large or too small for the existence of
quantum Hall phases. Although its origin is
different, this condition agrees with the result of \ste \ (at least for the
vortex quantum Hall phase).

\subsec{Periodic Chern-Simons terms and phase structure analysis}
In the following we shall analyze how the above picture is modified when we
impose the distinctive feature of Josephson junction arrays, namely the
periodicity of charge-vortex couplings, encoded in our formalism in the
Chern-Simons terms. This is achieved by introducing appropriate topological
excitations in the Euclidean lattice partition functions of \hidu and
\fihidu .

Let us begin with the gauge theory \hidu \ for the charge quantum Hall phase.
Its Euclidean lattice partition function coincides with \frm \ upon
substituting $e\to e'$, $B_{\mu }\to B_{\mu }^q = B_{\mu }+\nu _q A_{\mu }$
and rescaling the mixed Chern-Simons term by a factor of 2.
Therefore we present
here only the coupling of the topological excitations enforcing the
periodicity of the mixed Chern-Simons term $A_{\mu }k_{\mu \nu }B_{\nu }^q$
and the Wilson and 't Hooft loops \opa :
\eqn\newtopq{S_q = \sum _x \dots + ilp\sqrt{\kappa } A_{\mu }
\left( Q_{\mu }+M_{\mu } \right) + iln\sqrt{\kappa } B_{\mu }M_{\mu }
+ il {q\over {{\kappa }^{1\over 2}}} A_{\mu }q_{\mu }
+ il \phi \kappa ^{3\over 2} B_{\mu } \phi _{\mu } \ , }
where we have used the representation $\nu _q = p/n$.
Due to the change $B_{\mu }\to B_{\mu }^q$ the periodicities of the two
original gauge fields are changed from \shi \ to
\eqn\newshiq{\eqalign{A_{\mu } &\to A_{\mu } + {\pi n\over l\sqrt{\kappa }}
\ a_{\mu } \ , \qquad \qquad a_{\mu } \in Z\ ,\cr
B_{\mu } &\to B_{\mu } + {\pi p\over l\sqrt{\kappa }} \ b_{\mu } \ ,\qquad
\qquad b_{\mu }\in Z \ .\cr }}
The displayed terms in \newtopq \ can be rearranged as follows:
\eqn\rarrq{S_q = \sum _x \dots +ilp\sqrt{\kappa } A_{\mu } \left( Q_{\mu } +
{q\over \kappa p}q_{\mu }-{\kappa \phi \over n}\phi _{\mu } \right)
+ iln\sqrt{\kappa } B_{\mu }^q \left(
M_{\mu }+{\kappa \phi \over n} \phi _{\mu } \right) \ ,}
so that the whole model is reformulated in terms of the gauge
fields $A_{\mu }$
and $B_{\mu }^q$ and we can use thus the results of the previous section.
In particular we obtain
\eqn\whlq{\langle L_W L_H \rangle = {Z_{\rm Top}^q \left( Q_{\mu } +
{q\over \kappa p} q_{\mu }- {\kappa \phi \over n}\phi _{\mu }, M_{\mu }
+{\kappa \phi \over n} \phi _{\mu }
\right) \over Z_{\rm Top}^q \left( Q_{\mu }, M_{\mu } \right) } \ ,}
with
\eqn\topq{\eqalign{Z_{\rm Top}^q &= \sum _{\{ Q_{\mu } \}
\atop \{ M_{\mu } \} }
{\rm exp}\left( -S_{\rm Top}^q \right) \ ,\cr
S_{\rm Top}^q &= \sum_x {e'^2p^2\kappa \over 2l} \ Q_{\mu }
{\delta _{\mu \nu }
\over M_q^2-\nabla ^2 } Q_{\nu } + {g^2 n^2\kappa \over 2l}\ M_{\mu }
{\delta _{\mu \nu }\over M_q^2- \nabla ^2} M_{\nu } \cr
&\ \ \ \ \ \ \ \ \ + i{\pi pn M_q^2 \over l}
\ Q_{\mu }{k_{\mu \nu }\over \nabla ^2 \left( M_q^2-\nabla ^2 \right) }
M_{\nu } \ ,\cr }}
and $M_q$ defined in \masq .

At this point we can repeat verbatim the analysis of section 4. The phase
structure of \hidu \ with periodic Chern-Simons term is governed by the
topological excitations $Q_{\mu }$ and $M_{\mu }$. The phase in which both
these topological excitations are dilute corresponds to the charge
quantum Hall
phase discussed above, which constitutes an {\it incompressible fluid} of
charge-flux composites with short-range interactions.
The stability of this phase depends entirely
on the condensation conditions for the two types of topological excitations.
If $Q_{\mu }$ condenses we obtain a phase in which charges $q=\kappa p$
(and multiples thereof) are completely screened, while fluxes $\phi $ are
logarithmically confined: this is a conventional {\it superconducting phase}
with a charge $\kappa p$ condensate. Using the trick \nint \ we can identify
the Coulomb law for fluxes with the London form of the electromagnetic
response. If $M_{\mu }$ condenses we obtain, instead,
a phase in which excitations with quantum numbers $(q/(p\kappa )
-\phi \kappa /n)$ interact logarithmically. This means that the only
non-confined excitations in the model must carry both charge and flux
in the combination
\eqn\cvcom{{q\over \kappa p}- {\phi \kappa \over n}=0\  \Rightarrow
{q\over \phi } = \kappa ^2 \nu _q \ .}
These excitations are completely screened, while all other combinations
of quantum numbers are logarithmically confined. This (logarithmic)
{\it oblique confinement} \hoo \ \car \ phase describes thus
a {\it charge-flux superfluid} phase. Since the condensed composites carry
charge, this is actually
an {\it anyon superconductivity phase} \bobl
\ with a charge $p\kappa $ and vorticity $n/\kappa $ condensate. Indeed, the
electromagnetic response, obtained again by the trick \nint \ has still the
London form.

We now have to study the range of parameters in which these three phases are
realized. The analysis goes exactly as in section 4, giving the result
\eqn\psmgtq{\eqalign{X_q< 1 &\to \cases{{e'\over g}<{1\over \nu _q}\ ,
& conventional superconductor\ ,\cr {e'\over g} >{1\over \nu _q}\ ,
& anyon superconductor \ ,\cr }
\cr
X_q > 1 &\to  \cases{{e'\over g} < {1\over \nu _q X_q} \ ,
& conventional superconductor\ ,\cr
{1\over \nu _q X_q}<{e'\over g}<{X_q\over \nu _q} \ ,
& charge quantum Hall phase\ ,\cr
{e'\over g} > {X_q\over \nu _q} \ , & anyon superconductor \ ,\cr } \cr
X_q &\equiv {M_ql G\left(M_ql\right) \pi \over \mu } \ {pn\over 2}
\ .\cr }}
It is now harder to disentangle the phase diagram in terms of the original
(Josephson junction) parameters $e$ and $g$ since $e'$, and consequently $M_q$
depend themselves on the ratio $e/g$.

As was pointed out in \ste , the periodicity of charge-vortex
couplings is the
distinctive feature of Josephson junction arrays which allows
effective filling
fractions of order $O(1)$ and which is therefore expected to favour the
formation of charge and vortex quantum Hall phases. From the above result,
however, it is clear that the same mechanism can also destabilize these
quantum Hall phases as follows.
The condition for a charge quantum Hall phase is given by $X_q >1$.
The filling fraction parameters
$p$ and $n$ enter this condition in two ways. First there is the explicit
dependence of $X_q$ on the product $pn$; secondly there is the dependence of
the gap $M_q$ on the ratio $\nu_q = p/n$. The former favours
filling fractions
with a large product $pn$; too large numerators and denominators are however
presumably suppressed by the same mechanism as in the conventional quantum
Hall effect. The latter has the following effect. Decreasing both $\nu _q$
and $e/g$ makes both the gap $M_q$ and the ratio $e'/g$ smaller;
the parameter
$X_q$ approaches a fixed value, while $e'\nu _q/g$ can decrease indefinitely
till it becomes favourable for the system to expel the magnetic flux and form
a conventional superconductor. Increasing both $\nu _q$ and $e/g$, instead,
makes both the gap and the ratio $e'/g$ larger; for large values of
$M_ql$ the parameter $X_q$ tends to zero (see \met ), while $e'\nu _q/g$
can grow indefinitely and it quickly becomes favourable for the charge-flux
fluid to condense into a superfluid, so that the system becomes an anyon
superconductor. For values $e\nu _q/g>1$ we would expect a
charge-flux cristal.

The analysis of the gauge theory \fihidu \ for the vortex quantum Hall phase
follows exactly the same steps with all "electric quantities" and
"magnetic quantities" interchanged. Therefore we present here only the
final result:
\eqn\psmgtfi{\eqalign{X_{\phi }< 1 &\to \cases{{e\over g'}<{\nu _{\phi }}\ ,
& anyon superconductor\ ,\cr {e\over g'} >{\nu _{\phi }}\ ,
& superinsulator \ ,\cr } \cr
X_{\phi } > 1 &\to  \cases{{e\over g'} < {\nu _{\phi }\over X_{\phi }} \ ,
& anyon superconductor\ ,\cr
{\nu _{\phi }\over X_{\phi }}<{e\over g'}< \nu _{\phi }X_{\phi } \ ,
& vortex quantum Hall phase\ ,\cr
{e\over g'} > \nu _{\phi } X_{\phi } \ , & superinsulator \ ,\cr } \cr
X_{\phi } &\equiv {M_{\phi }l G\left(M_{\phi }l\right) \pi \over \mu }
\ {pn\over 2}
\ ,\cr }}
where $M_{\phi }$ is defined in \masfi \ and we have used the representation
$\nu _{\phi }=p/n$. Starting from the vortex quantum Hall phase and
decreasing both $g/e$ and $\nu _{\phi }$
makes both the gap $M_{\phi }$ and the ratio $g'/e$ smaller: it becomes
eventually favourable for the system to expel the offset charges and become
a superinsulator. In real Josephson junction arrays this phase
would presumably
be an insulating Abrikosov-type cristal of charges due to the
small but finite
ground capacitance $C_0$. Increasing both $g/e$ and $\nu _{\phi }$
makes both
the gap $M_{\phi }$ and the ratio $g'/e$ larger; the quantity $1/X_{\phi }$
tends to infinity (see \met ) while $e/g'$ decreases and the
flux-charge fluid
of the vortex quantum Hall phase condenses again into a flux-charge
superfluid, becoming thus an anyon superconductor. For even smaller
values of $e/g < \nu _{\phi }$ we would expect again a flux-charge cristal.

\newsec{Three dimensions}
In this section we generalize our results (for zero offset charges and
external
magnetic fluxes) to three-dimensional Josephson junction arrays.
While these are not (yet) experimentally accessible, we find it nonetheless
interesting
to construct their gauge theory representation and to study the differences
with the two-dimensional case. Clearly, self-duality is lost in three
dimensions since the fluctuating degrees of freedom are charges and closed
vortex loops. However, it is still possible to introduce the approximation
which allows a coupled gauge theory representation, as we now show.

Up to eq. \dif \ the analysis parallels exactly the two-dimensional case.
The decomposition analogous to \dif , however, requires the introduction
of the three-dimensional generalizations of the lattice
operators $K_{\mu \nu }$ and $\hat K_{\mu \nu }$. These are given
by the three-index lattice operators
\eqn\diaop{\eqalign{K_{\mu \nu \rho}
&\equiv S_{\mu }\epsilon _{\mu \alpha
\nu \rho } \Delta _{\alpha } \ ,\cr
\hat K_{\mu \nu \rho} &\equiv \epsilon _{\mu \nu \alpha \rho}
\hat \Delta _{\alpha }\hat S_{\rho } \ ,\cr }}
where $S_{\mu }$ and $\hat S_{\mu }$ are the shift operators
\dso \ and $\Delta _{\mu }$
and $\hat \Delta _{\mu }$ are the
finite difference operators \fdo . As in the two-dimensional case, these
two operators are interchanged (no minus sign) upon summation by parts
on the lattice. Moreover they are gauge invariant, in the sense that they
obey the following equations:
\eqn\gidiaop{\eqalign{K_{\mu \nu \rho} \Delta _{\nu } &= K_{\mu \nu \rho}
\Delta _{\rho } = \hat \Delta _{\mu } K_{\mu \nu \rho} = 0 \ ,\cr
\hat K_{\mu \nu \rho }\Delta _{\rho } &= \hat \Delta _{\mu }
\hat K_{\mu \nu \rho} = \hat \Delta _{\nu }
\hat K_{\mu \nu \rho } = 0 \ . \cr }}
Finally they satisfy also the equations
\eqn\prodiaop{\eqalign{\hat K_{\mu \nu \rho} K_{\rho \lambda \omega} &=
-\left( \delta _{\mu \lambda} \delta_{\nu \omega} - \delta _{\mu \omega}
\delta_{\nu \lambda } \right) \Delta  + \left( \delta _{\mu \lambda }
\Delta _{\nu } \hat \Delta _{\omega} - \delta _{\nu \lambda } \Delta _{\mu }
\hat \Delta _{\omega } \right) \cr
&\ \ \ + \left( \delta _{\nu \omega} \Delta _{\mu }
\hat \Delta _{\lambda } - \delta _{\mu \omega} \Delta _{\nu } \hat
\Delta _{\lambda } \right) \ ,\cr
\hat K_{\mu \nu \rho} K_{\rho \nu \omega } &= K_{\mu \nu \rho } \hat
K_{\rho \nu \omega} = 2 \left( \delta _{\mu \omega } \Delta - \Delta _{\mu }
\hat \Delta _{\omega } \right) \ .\cr }}

Using these operators we can decompose $v_{\mu }$ as
\eqn\tdif{v_{\mu } = \Delta _{\mu }m + \Delta _{\mu } \alpha +
K_{\mu \alpha \beta }\psi _{\alpha \beta } \ ,}
with $m\in Z$ and $|\alpha |<1$. The $\psi _{\alpha \beta}$'s are
restricted
by the fact that the antisymmetric combinations $q_{\mu \nu } \equiv
\hat K_{\mu \nu \alpha }v_{\alpha }= \hat K_{\mu \nu \alpha }
K_{\alpha \lambda
\rho } \psi _{\lambda \rho }$ must be integers. We can thus trade
the original
sum over the four independent integers $\{ v_{\mu } \}$ for a sum over the
seven integers $\{ m, q_{\mu \nu } \}$ subject to the constraint
$\hat \Delta _{\mu }q_{\mu \nu } = \hat \Delta _{\nu } q_{\mu \nu } =0$.
This constraint eliminates the three longitudinal degrees of freedom of
$q_{\mu \nu }$, so that $\{ m, q_{\mu \nu } \}$ with the above constraint
describes only four independent integers. After shifting the
$\Phi $ integration domain using the sum over $\{ m \}$ and performing the
resulting trivial $\Phi $ integration we are left with
\eqn\tpfjjd{\eqalign{Z &= \sum _{\{ q_{\mu \nu} \}} \delta_{\hat \Delta_{\mu }
q_{\mu \nu}, 0} \ \int {\cal D}p_{\mu } \ \delta \left(
\hat \Delta _{\mu }p_{\mu } \right) \ {\rm exp }(-S)\ ,\cr
S &= \sum_x -i2\pi \ p_{\mu } K_{\mu \alpha \beta} \psi_{\alpha \beta }
+ N^2 l_0 E_C \ p_0 {1\over -\Delta } p_0 + {p_i^2 \over 2l_0E_J} \ .\cr }}

The constraints are solved by introducing a real {\it antisymmetric
gauge field } $b_{\mu \nu }$ and an integer gauge field $a_{\mu }$:
\eqn\tsco{\eqalign{ p_{\mu } &\equiv K_{\mu \alpha \beta }b_{\alpha \beta }
\ , \qquad \qquad b_{\alpha \beta }\in R\ ,\cr
q_{\mu \nu } &\equiv \hat K_{\mu \nu \alpha} a_{\alpha }\ ,\qquad
\qquad a_{\alpha }\in Z\ .\cr }}
Repeating the same steps as in the two-dimensional case we find
\eqn\tpfjje{\eqalign{Z &= \sum _{\{ Q_{\mu } \}} \int {\cal D}a_{\mu }
\int {\cal D}b_{\mu } \ {\rm exp}(-S) \ ,\cr
S &= \sum_x -i2\pi \ a_{\mu }K_{\mu \alpha \beta }b_{\alpha \beta }
+ N^2l_0E_C \ p_0 {1\over -\Delta }p_0 + {p_i^2\over 2l_0E_J}
+ i2\pi a_{\mu }Q_{\mu } \ ,\cr }}
which is the three-dimensional analogue of \pfjje . Here, $K_{\mu \alpha
\beta }b_{\alpha \beta }$ maintains its interpretation of the conserved
four-current of charge fluctuations, while
$\hat K_{\mu \nu \alpha }a_{\alpha}$
represents the fluctuations of {\it closed vortex loops}.

Since the magnetic fluctuations are represented by closed vortex loops we
cannot render the partition function self-dual, as in the two-dimensional
case. However, it is still possible to introduce a bare kinetic term for
the vortex loops with a coefficient tuned so that it can be combined with
the Coulomb term for the charges into $\sum_x {\pi ^2 \over 2N^2 l_0 E_C}
q_{ij}q_{ij} $. As in two dimensions we have to introduce new integers via
the Poisson summation formula to guarantee that the charge current
$K_{\mu \alpha \beta} b_{\alpha \beta }$ remains an integer. In three
dimensions these are two-index antisymmetric integers $M_{\mu \nu }$
satisfying the constraint $\hat \Delta _{\mu } M_{\mu \nu } = \hat \Delta
_{\nu } M_{\mu \nu } =0$:
\eqn\tsda{\eqalign{Z &= \sum_{\{ Q_{\mu } \} \atop \{ M_{\mu \nu} \} }
\int {\cal D}a_{\mu } \int {\cal D}b_{\mu } \ {\rm exp}(-S)\ ,\cr
S &= \sum_x -i2\pi \ a_{\mu }K_{\mu \alpha \beta }b_{\alpha \beta }
+ {p_i^2\over 2l_0E_J}
+{\pi ^2 q_{ij}^2 \over 2N^2l_0E_C} + i2\pi a_{\mu }Q_{\mu } +i 2\pi b_{\mu }
M_{\mu \nu} \ .\cr }}
After a rescaling
\eqn\tres{\eqalign{A_0 &\equiv {2\pi \over \sqrt{N}l_0} a_0\ ,\qquad \qquad
A_i \equiv {2\pi \over \sqrt{N}l} a_i \ ,\cr
B_{0i} &\equiv {2\pi \over \sqrt{N}l_0l} b_{0i} \ ,\qquad \qquad
B_{ij} \equiv {2\pi \over \sqrt{N}l^2} b_{ij} \ ,\cr }}
we obtain finally
\eqn\tsdb{\eqalign{Z &= \sum_{ \{ Q_{\mu } \} \atop \{ M_{\mu \nu} \} }
\int {\cal D}A_{\mu } \int {\cal D}B_{\mu } \ {\rm exp} (-S) \ ,\cr
S &= \sum_x {l_0 l^3\over 4e^2}
\ \tilde F_{ij}\tilde F_{ij} -i {l_0l^3 \kappa \over 2\pi }
A_{\mu }k_{\mu \alpha \beta }B_{\alpha \beta } + {l_0l^3 \over 2g^2}
\ f_if_i \cr
&\ \ \ \ \ \ \ \ + i\sqrt{\kappa}
\left( l_0Q_0A_0 + lQ_iA_i \right) +i \sqrt{\kappa}
\left( l_0l M_{0i} B_{0i} + l_0l M_{i0} B_{i0}
+ l^2 M_{ij} B_{ij} \right) \ ,\cr }}
where $k_{\mu \nu \rho }$ and $\hat k_{\mu \nu \rho}$ are the analogues of
\diaop \ defined in terms of derivative operators rather than finite
difference operators (and satisfying the correspondingly modified
eq. \prodiaop ) , $\tilde F_{ij}$ and $f_i$ are the spatial components of
\eqn\tdfstq{\eqalign{\tilde F_{\mu \nu } &\equiv \hat k_{\mu \nu \alpha }
A_{\alpha } \ ,\cr
f_{\mu } &\equiv  {1\over 2} k_{\mu \alpha \beta } B_{\alpha \beta } \ ,\cr }}
and the coupling constants $e^2$ (dimensionless) and $g^2$ (with dimensions
${\rm mass}^2$) are given by
\eqn\tide{e^2 = 2N lE_C\ , \qquad \kappa = N\ , \qquad g^2={\pi ^2\over
N l} E_J \ .}
The plasma frequency is given again by a product of these coupling constants,
\eqn\tplf{m= {egN\over \pi } = \sqrt{ 2N^2 E_CE_J} \ ,}
and for $ml \le O(1)$ we can choose $l_0=l$. Also in three dimensions we have
thus obtained a coupled gauge theory in the limit of infinite magnetic
permeabilities, encoded in the absence of the time components of both
$\tilde F_{\mu \nu }$ and $f_{\mu }$ in the kinetic terms. As in two
dimensions we shall henceforth consider the relativistic limit of this
gauge theory:
\eqn\tfrm{\eqalign{Z &= \sum _{ \{ Q_{\mu } \} \atop \{ M_{\mu \nu} \} }
\int {\cal D}A_{\mu } \int {\cal D}B_{\mu } \ {\rm exp} (-S) \ ,\cr
S &= \sum_x {l^4\over 4e^2}
\tilde F_{\mu \nu}\tilde F_{\mu \nu} -i{l^4\kappa \over 2\pi }
A_{\mu }k_{\mu \alpha \beta }B_{\alpha \beta } + {l^4\over 2g^2}
f_{\mu }f_{\mu } +il\sqrt{\kappa } A_{\mu }Q_{\mu }
+il^2\sqrt{\kappa } B_{\mu \nu}M_{\mu \nu} \ .\cr }}
This is a pure gauge theory representation
of a model of interacting charges and closed vortex loops.
The identification of the physical degrees of freedom is analogous to the
two-dimensional case:
\eqn\tcav{\eqalign{q_{\mu }^{\rm charge} &\equiv {{\kappa }^{3\over 2}\over
2\pi } \ k_{\mu \alpha \beta }B_{\alpha \beta } \ ,\cr
\phi _{\mu \nu}^{\rm vortex} &\equiv {1\over 2\pi {\kappa}^{1\over 2}}
\ \hat k_{\mu \nu \alpha }A_{\alpha } \ .}}

The model \tfrm \ is a (Euclidean) lattice version of the so called {\it BF
gauge theory} \bal , whose Lagrangian is given by
\eqn\lbfgt{\eqalign{{\cal L}_{BF} &= -{1\over 12g^2} f_{\mu \nu \rho}
f^{\mu \nu \rho } + {\kappa \over 4\pi } B_{\mu \nu } \epsilon ^{\mu \nu
\lambda \rho } F_{\lambda \rho }- {1\over 4e^2} F_{\mu \nu }
F^{\mu \nu } \ ,\cr
f_{\mu \nu \rho} &\equiv \partial _{\mu }B_{\nu \rho} + \partial _{\nu }
B_{\rho \mu } + \partial _{\rho } B_{\mu \nu } \ ,\cr
f^{\mu } &\equiv {1\over 6} \epsilon ^{\mu \nu \lambda \rho } f_{\nu \lambda
\rho } \ .\cr }}
Here $A_{\mu }$ describes an ordinary (3+1)-dimensional photon with
field strength given by $F_{\mu \nu} = \partial _{\mu }A_{\nu }-
\partial _{\nu } A_{\mu }$ and dual field strength $\tilde F^{\mu \nu } =
{1\over 2} \epsilon
^{\mu \nu \lambda \rho }F_{\lambda \rho }$. This photon has a topological
$BF$ coupling to an antisymmetric Kalb-Ramond \kal \ gauge field
$B_{\mu \nu}$,
whose field strength is given by the three-form $F_{\mu \nu \rho}$.
The first
term in \lbfgt \ represents the kinetic term for the
Kalb-Ramond gauge field.
In addition to the usual invariance under gauge transformations
of $A_{\mu }$
\lbfgt \ is also invariant under gauge transformations
$B_{\mu \nu } \to B_{\mu \nu }+\partial _{\mu } \Lambda _{\nu }
- \partial _{\nu }\Lambda _{\mu }$. The Kalb-Ramond gauge theory
describes a single
massless scalar degree of freedom since all the components $B_{i0}$ are
Lagrange multipliers and $B_{ij}$ has only one transverse component.
When coupled to the usual Maxwell theory via a $BF$ term, the Kalb-Ramond
sector induces a topological mass
\eqn\krmass{m={eg\kappa \over \pi }}
for the photon. As already pointed out in \tplf \ this mass represents the
plasma frequency of Josephson junction arrays.
The $BF$ system is thus the natural three-dimensional
generalization of \mcs , as expected.

As in two dimensions, the integer-valued variables $Q_{\mu }$
and $M_{\mu \nu }$ appearing in \tfrm \ represent (Euclidean) topological
excitations whose
role is to make the charge-vortex $BF$ coupling {\it periodic}. They satisfy
the constraints
\eqn\tconst{\eqalign{\hat d_{\mu } Q_{\mu } &=0\ ,\cr
\hat d_{\mu } M_{\mu \nu } &= \hat d_{\nu } M_{\mu \nu } =0\ .\cr }}
The "electric" topological excitations $Q_{\mu }$ are exactly as in two
dimensions; the "magnetic" topological excitations, instead, describe
{\it compact surfaces} on the lattice (in a dilute phase) or
infinite surfaces (in a dense phase).

The phase structure of three-dimensional Josephson junction arrays (in our
approximation) is thus determined by the statistical mechanics of a coupled
gas of lattice loops and surfaces. Its partition function can be easily
obtained by a Gaussian integration over the gauge fields $A_{\mu }$ and
$B_{\mu \nu }$ in \tfrm :
\eqn\ttop{\eqalign{Z_{\rm Top} &=
\sum_{ \{ Q_{\mu } \} \atop \{ M_{\mu \nu} \} }
\ {\rm exp}\left( -S_{\rm Top} \right) \ ,\cr
S_{\rm Top} &= \sum_x {e^2\kappa \over 2l^2} \ Q_{\mu }{\delta _{\mu \nu }
\over {m^2- \nabla ^2 }} Q_{\nu } + {g^2 \kappa \over 2} \ M_{\mu \nu}
{{\delta _{\mu \alpha } \delta _{\nu \beta }-\delta _{\mu \beta}
\delta _{\nu \alpha }} \over {m^2-\nabla ^2}} M_{\alpha \beta } \cr
&\ \ \ \ \ \ \ \ +i {\pi m^2 \over l}
\ Q_{\mu } {k_{\mu \alpha \beta}
\over {\nabla ^2 \left( m^2- \nabla ^2 \right) }} M_{\alpha \beta } \ ,\cr }}
with $m$ defined in \krmass .
This partition function can be interpreted as the Euclidean partition
function for a lattice model of interacting particles (whose world-lines are
parametrized by the closed loops $Q_{\mu }$) and closed Nielsen-Olesen type
strings \froc \ (whose world-sheets are parametrized by the compact surfaces
$M_{\mu \nu }$). In a derivative expansion the string action takes the form
\eqn\sadex{S_{\rm strings} = \sum_x {\pi ^2 \over \kappa e^2} M_{\mu \nu }
M_{\mu \nu } + \dots \ .}
In the dilute gas approximation, where $M_{\mu \nu}$ can take only the
values $0, \pm 1$,
this term measures the {\it area} of the world-sheet and is thus the
standard Nambu-Goto term \pol , with {\it string tension}
$\pi ^2 /\kappa e^2 l^2$ (remember that $1/e^2l^2$ was the vortex mass in two
dimensions). The parameter $1/l$ plays thus the role of the Higgs mass in our
lattice model; higher order terms in \sadex \ involve both the curvature and
internal excitations of the string.
Particle-string interactions are encoded in the topological Aharonov-Bohm
term (third term in \ttop ) measuring the linking of the closed
world-lines of
particles and compact world-sheets of strings in four Euclidean dimensions.
As in the two-dimensional case this term vanishes for loops and surfaces
separated on distances much larger than $1/m$. In this case the denominator
reduces to $m^2\nabla ^2$ and, by using either one of the representations
\eqn\rclsu{\eqalign{Q_{\mu } &= lk_{\mu \alpha \beta} Y_{\alpha \beta}\ ,\cr
M_{\mu \nu } &= l\hat k_{\mu \nu \alpha } X_{\alpha }\ ,\cr }}
and the equations \prodiaop \ one recognizes that the whole term reduces
to ($i2\pi {\rm integer}$), which is equivalent to 0.

Unfortunately, the statistical mechanics of random surfaces \froc \ is much
less understood than its random loop counterpart \loo \ and we cannot use
self-duality arguments anymore. In order to proceed further we shall assume
that the same three types of phases as in the two-dimensional case can exists
and we will point out the differences that can nonetheless arise. First of
all we can repeat the same argument as in the two-dimensional case to find
that in the electric condensation phase the global gauge symmetry associated
with $A_{\mu }$ is spontaneously broken $R_A\to Z_A$ while in the magnetic
condensation phase it is the Kalb-Ramond global gauge symmetry which is
spontaneously broken $R_B\to Z_B$. Secondly we can consider
the expectation values of the Wilson loop \opa \ and
the {\it 't Hooft surface}
\eqn\hoosur{S_H \equiv {\rm exp} \left( i\phi \kappa ^{3\over 2} \sum_x
l^2 \phi_{\mu \nu } B_{\mu \nu } \right) \ ,}
where $\phi _{\mu \nu }$ vanishes everywhere but on the plaquettes of a
compact surface, where it takes the value 1.These expectation values are
given by
\eqn\evwh{\eqalign{\langle L_W \rangle &= {{Z_{\rm Top} \left( Q_{\mu } +
{q\over \kappa } q_{\mu }, M_{\mu \nu} \right) } \over {Z_{\rm Top} \left(
Q_{\mu }, M_{\mu \nu} \right) }} \ ,\cr
\langle S_H \rangle &= {{Z_{\rm Top } \left( Q_{\mu }, M_{\mu \nu}
+ \phi \kappa
\phi _{\mu \nu} \right) } \over {Z_{\rm Top} \left( Q_{\mu },
M_{\mu \nu} \right) }} \ .\cr }}
With exactly the same computation as in the two-dimensional case we find
the dominant contributions
\eqn\tdomcon{\eqalign{\langle L_W \rangle _{\rm mag.\ cond.} &=
{\rm exp} \left( {-e^2 q^2 \over 2\kappa l^2} q_{\mu } {\delta _{\mu \nu }
\over -\nabla ^2} q_{\nu } \right) \ ,\cr
\langle S_H \rangle _{\rm el.\ cond.} &= {\rm exp} \left( -{g^2 \kappa ^3
\phi ^2 \over 2} \phi _{\mu \nu } {{\delta _{\mu \alpha }\delta _{\nu \beta}
-\delta _{\mu \beta}\delta _{\nu \alpha }}\over -\nabla ^2} \phi _{\alpha
\beta } \right) \ .\cr }}
Small loop (surface) corrections can alter only the coefficients of
the Coulomb potentials in these results. The long-range nature of the
interaction kernels is associated with the Goldstone bosons due to the
spontaneous symmetry breaking present in both the electric and the
magnetic condensation phases.
In four Euclidean dimensions the results \tdomcon \ for the 't Hooft
surface
in the electric condensation phase implies that the self-energy of
a circular
vortex loop of radius $R$ is proportional to $R {\rm ln}R$. As in two
dimensions this is tantamount to {\it logarithmic confinement} of magnetic
fluxes and we conclude thus that the electric condensation phase is still a
{\it superconducting phase}. The result \tdomcon \ for the Wilson loop in
the magnetic condensation phase, instead, represents a {\it perimeter law},
implying a $1/r$ Coulomb potential between charges. The amount
of energy required to separate a charge-anticharge pair is finite, although
the interaction potential is long-range. We identify this as an
{\it insulating phase} (as opposed to the superinsulator in two dimensions).
Clearly, the dilute phase for both topological excitations corresponds
again to a {\it metallic phase}. As already mentioned, it is
harder to estimate
the position of the phase transitions in three dimensions due to
the lack of
self-duality.

\listfigs
\listrefs
\end